\documentclass[twocolumn, journal]{IEEEtran}
\IEEEoverridecommandlockouts
\usepackage{color}
\usepackage{graphicx}
\usepackage{epstopdf}
\usepackage{amsmath}
\usepackage{amssymb}
\usepackage{algorithm}
\usepackage{algorithmic}
\usepackage{amsmath}
\usepackage{multirow}
\usepackage{booktabs}
\usepackage{array}
\usepackage{amsthm}
\usepackage{stfloats}
\usepackage{caption}
\usepackage{subfigure}
\usepackage{bm}
\usepackage{setspace}
\usepackage{diagbox}
\usepackage{flushend}
\usepackage{enumitem}

\newcommand{\be}{\begin{equation}}
\newcommand{\ee}{\end{equation}}
\newcommand{\bea}{\begin{eqnarray}}
\newcommand{\eea}{\end{eqnarray}}
\newcommand{\ba}{\begin{array}}
\newcommand{\ea}{\end{array}}

\allowdisplaybreaks[4]

\def\BibTeX{{\rm B\kern-.05em{\sc i\kern-.025em b}\kern-.08em
    T\kern-.1667em\lower.7ex\hbox{E}\kern-.125emX}}
\begin{document}

\title{Tri-timescale Beamforming Design for Tri-hybrid Architectures with Reconfigurable Antennas
\thanks{M. Liu and M. Li are with the School of Information and Communication Engineering, Dalian University of Technology, Dalian 116024, China (e-mail: liumengzhen@mail.dlut.edu.cn, mli@dlut.edu.cn).}
\thanks{R. Liu is with the Center for Pervasive Communications and Computing, University of California, Irvine, CA 92697, USA (e-mail: rangl2@uci.edu).}
\thanks{Q. Liu is with the School of Computer Science and Technology, Dalian University of Technology, Dalian 116024, China (e-mail: qianliu@dlut.edu.cn).}}

\author{Mengzhen Liu,~\IEEEmembership{Graduate Student Member,~IEEE,}
        Ming Li,~\IEEEmembership{Senior Member,~IEEE,}
        Rang Liu,~\IEEEmembership{Member,~IEEE,}\\
        and Qian Liu,~\IEEEmembership{Member,~IEEE}
\vspace{-0.0 cm}
}

\maketitle
\pagestyle{empty}  
\thispagestyle{empty} 

\begin{abstract}
Reconfigurable antennas possess the capability to dynamically adjust their fundamental operating characteristics, thereby enhancing system adaptability and performance. To fully exploit this flexibility in modern wireless communication systems, this paper considers a novel tri-hybrid beamforming architecture, which seamlessly integrates pattern-reconfigurable antennas with both analog and digital beamforming. The proposed tri-hybrid architecture operates across three layers: (\textit{i}) a radiation beamformer in the electromagnetic (EM) domain for dynamic pattern alignment, (\textit{ii}) an analog beamformer in the radio-frequency (RF) domain for array gain enhancement, and (\textit{iii}) a digital beamformer in the baseband (BB) domain for multi-user interference mitigation.
To establish a solid theoretical foundation, we first develop a comprehensive mathematical model for the tri-hybrid beamforming system and formulate the signal model for a multi-user multi-input single-output (MU-MISO) scenario. The optimization objective is to maximize the sum-rate while satisfying practical constraints. Given the challenges posed by high pilot overhead and computational complexity, we introduce an innovative tri-timescale beamforming framework, wherein the radiation beamformer is optimized over a long-timescale, the analog beamformer over a medium-timescale, and the digital beamformer over a short-timescale. This hierarchical strategy effectively balances performance and implementation feasibility. Simulation results validate the performance gains of the proposed tri-hybrid architecture and demonstrate that the tri-timescale design significantly reduces pilot overhead and computational complexity, highlighting its potential for future wireless communication systems, and revealing a practical way to exploit EM‑domain reconfigurability across three natural channel timescales.
\end{abstract}


\vspace{-0.3cm}
\section{Introduction}
Future wireless communication systems are anticipated to leverage ultra-wide spectrum resources in high-frequency bands, such as millimeter-wave (mmWave), to meet the escalating demand for ultra-high data rates \cite{W. Saad 2020 6G}. To counteract severe path loss in these bands, massive multiple-input multiple-output (massive MIMO) systems or extremely large antenna arrays (ELAA) have emerged as promising technologies to ensure reliable signal transmission \cite{E. Bjornson 2019 MIMO}. In conventional antenna arrays, each antenna element typically possesses a fixed radiation pattern, concentrating energy toward predetermined directions, often perpendicular to the array plane. Consequently, existing research primarily focuses on signal-processing-based beamforming methods that manipulate the phase and amplitude of antenna signals across antenna elements to steer beams toward specific directions.

However, the limitations of conventional antennas with fixed radiation patterns become increasingly evident in emerging applications, such as ELAA and near-field communications \cite{C. You 2025 NGAT}. Specifically, system performance critically depends not only on array-level beamforming but also on individual antenna element radiation characteristics. Narrow radiation patterns with high energy concentration can deliver strong gains within specific angular sectors but experience significant performance deterioration in end-fire directions due to sharp gain reductions. Conversely, antennas with broader radiation patterns provide extended coverage at the expense of lower gain, ultimately degrading user quality of service (QoS).

Reconfigurable antenna technology offers a compelling approach to addressing these limitations by introducing enhanced adaptability \cite{W. K. New 2025 tutorial}-\cite{H. Jafarkhani 2018 reconfigurable TWC}. Such antennas dynamically adjust their internal configurations, modifying key operational characteristics, including radiation patterns \cite{R. Murch 2022 pixel fig_reference_round}-\cite{R. Murch 2022 pixel pattern endfire}, frequency responses \cite{R. Murch 2014 frequency}, and polarizations \cite{R. W. Heath 2024 polarization}, \cite{H. Li 2024 polarization}. Among these, radiation pattern reconfiguration is particularly significant for wireless communication, directly influencing signal strength at transmission and reception. Several hardware implementations facilitating flexible control have emerged, including electronically steerable parasitic array radiator (ESPAR) antennas \cite{R. Murch 2023 ESPAR}, frequency selective surfaces (FSS) \cite{FSS}, and pixel antennas \cite{K.K.Wong 2025 pixel}. Typically, a reconfigurable antenna comprises an active radiating element and multiple parasitic elements, whose connection states are dynamically adjusted to alter current distribution and thereby modify electromagnetic (EM) properties. This adaptability provides new degrees of freedom (DoF) in the EM domain, offering substantial potential to overcome existing performance limitations and enable more efficient wireless communication systems.

Recent advances in reconfigurable antenna technology have sparked growing interest in integrating pattern-reconfigurable antennas into arrays. For instance, the authors in \cite{K.K.Wong R. Murch 2024 Antenna coding pixel} employ multi-port network theory and beamspace channel representation to develop an EM-based communication model for pixel antennas, which is then used to optimize antenna coding for maximizing channel gain.
Additionally, the authors in \cite{Z. Gao 2024 EM domain} introduce a decomposition method based on spherical harmonic functions to model radiation patterns in the EM domain. This approach decomposes the radiation pattern into a linear combination of orthogonal bases, facilitating the iterative design of radiation patterns and digital precoders. While these studies highlight the advantages of replacing conventional pattern-fixed antennas with innovative pattern-reconfigurable ones, they often neglect the growing concerns of hardware complexity and energy consumption that future wireless communication systems will encounter. Such concerns become increasingly critical in large-scale deployments, where both the antenna count and operating frequency bands continue to expand. To tackle these issues, existing literature commonly advocates hybrid beamforming architectures for mMIMO systems, as a power-efficient alternative to conventional fully-digital beamforming \cite{C. Han 2021}.

Hybrid beamforming architectures typically partition functionality between analog beamforming, controlling beam directions in the radio-frequency (RF) domain, and digital beamforming, performing spatial signal processing in the baseband (BB) domain. Utilizing fewer RF chains and hardware-efficient phase shifters (PSs), various hybrid architectures, such as fully-connected \cite{F. Sohrabi 2015}, partially-connected \cite{L. Dai 2016}, and dynamic structures \cite{H. Li 2020 Dynamic}, achieve a balanced trade-off between performance and complexity. However, these architectures inherently rely on antennas with fixed radiation characteristics, severely restricting their adaptability to dynamic propagation environments and limiting overall system performance.

To overcome this critical limitation, leveraging pattern-reconfigurable antennas within hybrid beamforming architectures emerges as an effective solution, enabling dynamic adaptation of radiation characteristics while maintaining reduced hardware complexity and power consumption. In this paper, we consider a novel tri-hybrid beamforming architecture that integrates pattern-reconfigurable antennas with both analog and digital beamforming. This architecture can be viewed as an advanced extension of existing reconfigurable antenna-based frameworks. Specifically, the proposed tri-hybrid architecture comprises three distinct layers: an EM domain radiation beamformer aligning radiation patterns dynamically, an RF domain analog beamformer enhancing array gains, and a BB domain digital beamformer mitigating multi-user interference.

The concept of this tri-hybrid architecture was first introduced in \cite{R. W. Heath 2023 tri-hybrid}-\cite{R. W. Heath 2025 tri-hybrid}. While prior analyses and simulations have demonstrated the advantages of tri-hybrid beamforming in terms of spectral efficiency and energy efficiency, several critical issues must be addressed before its practical implementation in future wireless systems. Firstly, acquiring instantaneous channel state information (CSI) becomes considerably more challenging due to increased dimensionality and limited RF chains, resulting in substantial pilot overhead. Secondly, jointly optimizing beamformers across three distinct domains introduces considerable complexity, primarily due to parameter coupling, discrete phase-shift constraints, and the nonlinear characteristics inherent to reconfigurable antennas.

To address these challenges, recent studies proposed a two-timescale beamforming framework to mitigate pilot overhead and computational complexity. Specifically, analog beamforming optimization leverages statistical CSI over a longer timescale, whereas digital beamforming employs instantaneous, low-dimensional effective CSI over a shorter timescale \cite{A. Liu two-stage}-\cite{A. Liu 2022 sparse}. Inspired by this approach and capitalizing on the distinctive properties of our tri-hybrid architecture, we introduce an innovative tri-timescale beamforming framework, effectively resolving practical implementation challenges, significantly reducing channel estimation overhead, and lowering computational complexity.
The main contributions of this paper are summarized as follows:

\begin{itemize} \item We propose a comprehensive mathematical model for the radiation beamformer based on multiple pattern-reconfigurable antennas, including the introduction of a novel virtual angular index channel in the EM domain. Utilizing this model, we develop a tri-hybrid beamforming architecture, formulate a detailed multi-user multiple-input single-output (MU-MISO) signal model, and rigorously establish the corresponding sum-rate maximization optimization problem under realistic system constraints.

\item To overcome the inherent challenges of excessive pilot overhead and high computational complexity, we introduce an innovative tri-timescale beamforming framework. Specifically, we optimize the radiation beamformer at a long timescale according to the statistical CSI of scattering cluster cores, update the analog beamformer at a medium timescale based on high-dimensional statistical CSI of multipath rays, and adaptively design the digital beamformer at a short timescale using instantaneous low-dimensional effective CSI.

\item We conduct extensive simulations to verify the effectiveness and scalability of the proposed tri-hybrid beamforming architecture. Particularly, the proposed tri-hybrid beamforming achieves around 7dB transmit power reduction at a given sum-rate and nearly an order-of-magnitude pilot overhead reduction compared to real-time beamforming, underscoring its practical advantages for next-generation wireless systems.

\end{itemize}

\begin{figure}[!t]
    \centering
    \vspace{-0.0cm}
    \subfigure{{\includegraphics[width= 1.3 in]{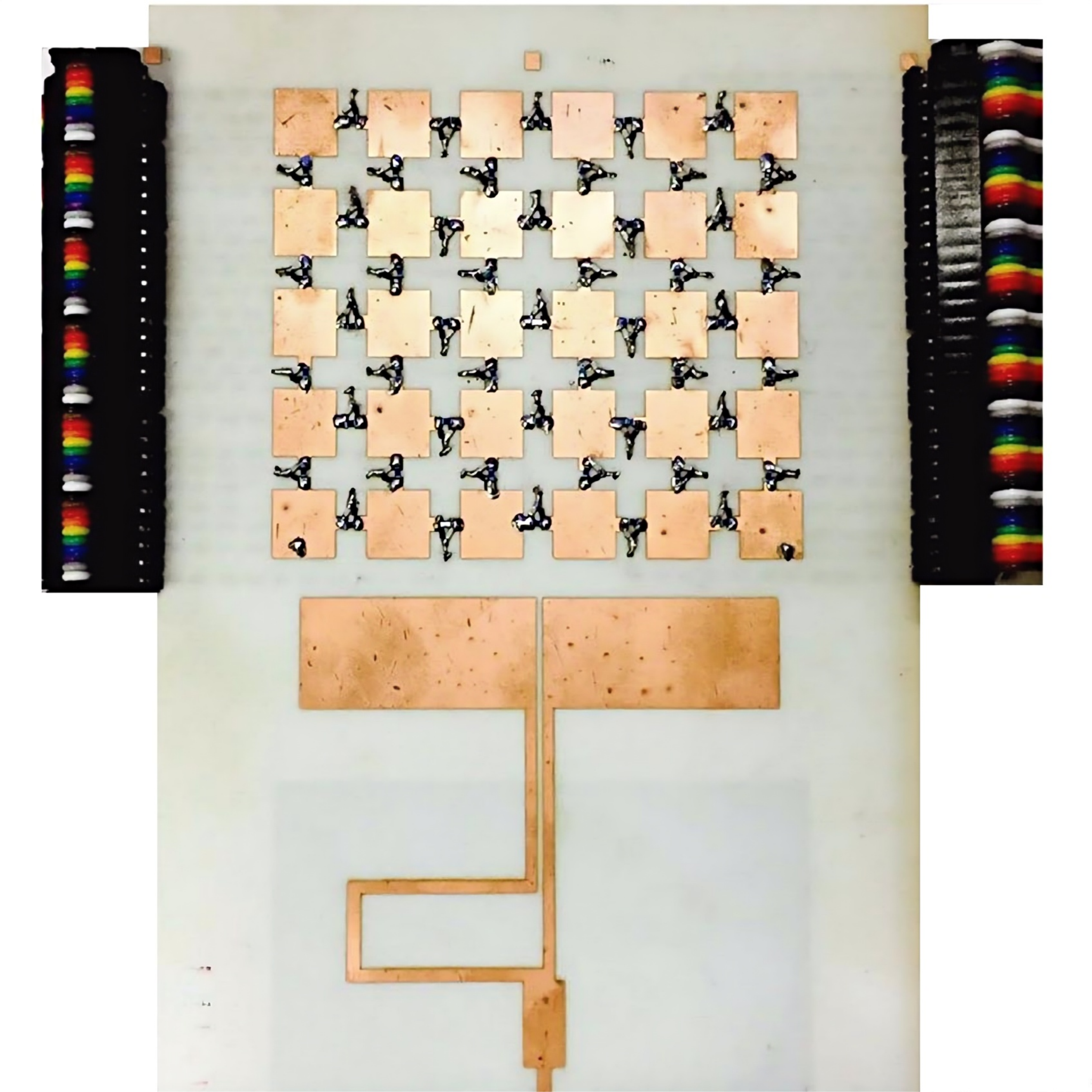}}}
    \hspace{0.3 cm}
    \subfigure{{\includegraphics[width= 1.3 in]{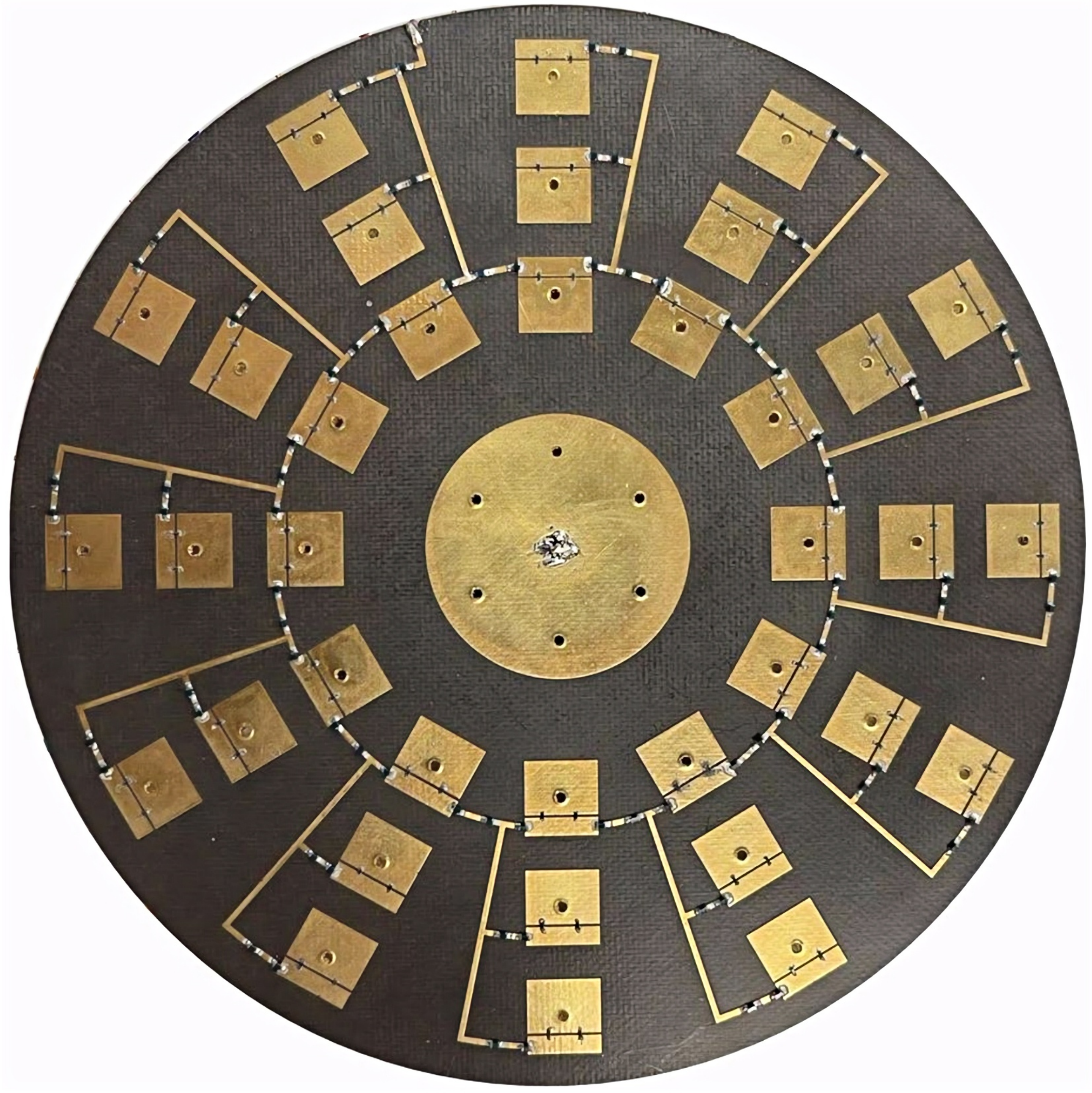}}}
    \caption{The hardware prototypes of pattern-reconfigurable antenna (left \cite{R. Murch 2017 pixel fig_reference_square}, right \cite{R. Murch 2022 pixel pattern endfire}).}
     \vspace{-0.4 cm}
    \label{fig:hardware}
\end{figure}

\textit{Notations}: Boldface lower-case and upper-case letters indicate column vectors and matrices, respectively. $(\cdot)^{*}$, $(\cdot)^{T}$, and $(\cdot)^{H}$ denote the conjugate, transpose, and transpose-conjugate. $|a|$, $\|\mathbf{a}\|$, and $\|\mathbf{A}\|_{F}$ are the magnitude of scalar $a$, the norm of vector $\mathbf{a}$, and the Frobenius norm of matrix $\mathbf{A}$. $\mathrm{vec}(\mathbf{A})$ stacks $\mathbf{A}$'s columns into a long column vector. Notation $\otimes$ is the Kronecker product of matrices. $\mathrm{blkdiag}\{\mathbf{A}_{1},\mathbf{A}_{2},\ldots,\mathbf{A}_{M}\}$ denotes a block diagonal matrix. $\mathbf{1}_{M}$ and $\mathbf{I}_{M}$ represent an $M\!\times\!1$ vector of ones and an $M\!\times\!M$ identity matrix, respectively. $\Re\{\cdot\}$ denotes the real part of a complex number. The statistical expectation is given by $\mathbb{E}\{\cdot\}$. $\mathbf{a}(i)$ denotes the $i$-th entry of the vector $\mathbf{a}$. $\mathbf{A}(i,:)$ and $\mathbf{A}(i,j)$ denote the $i$-th row and $(i,j)$-th element of matrix $\mathbf{A}$, respectively.

\section{System Model and Problem Formulation}

\begin{figure}[!t]
  \centering
  \includegraphics[width= 2.8 in]{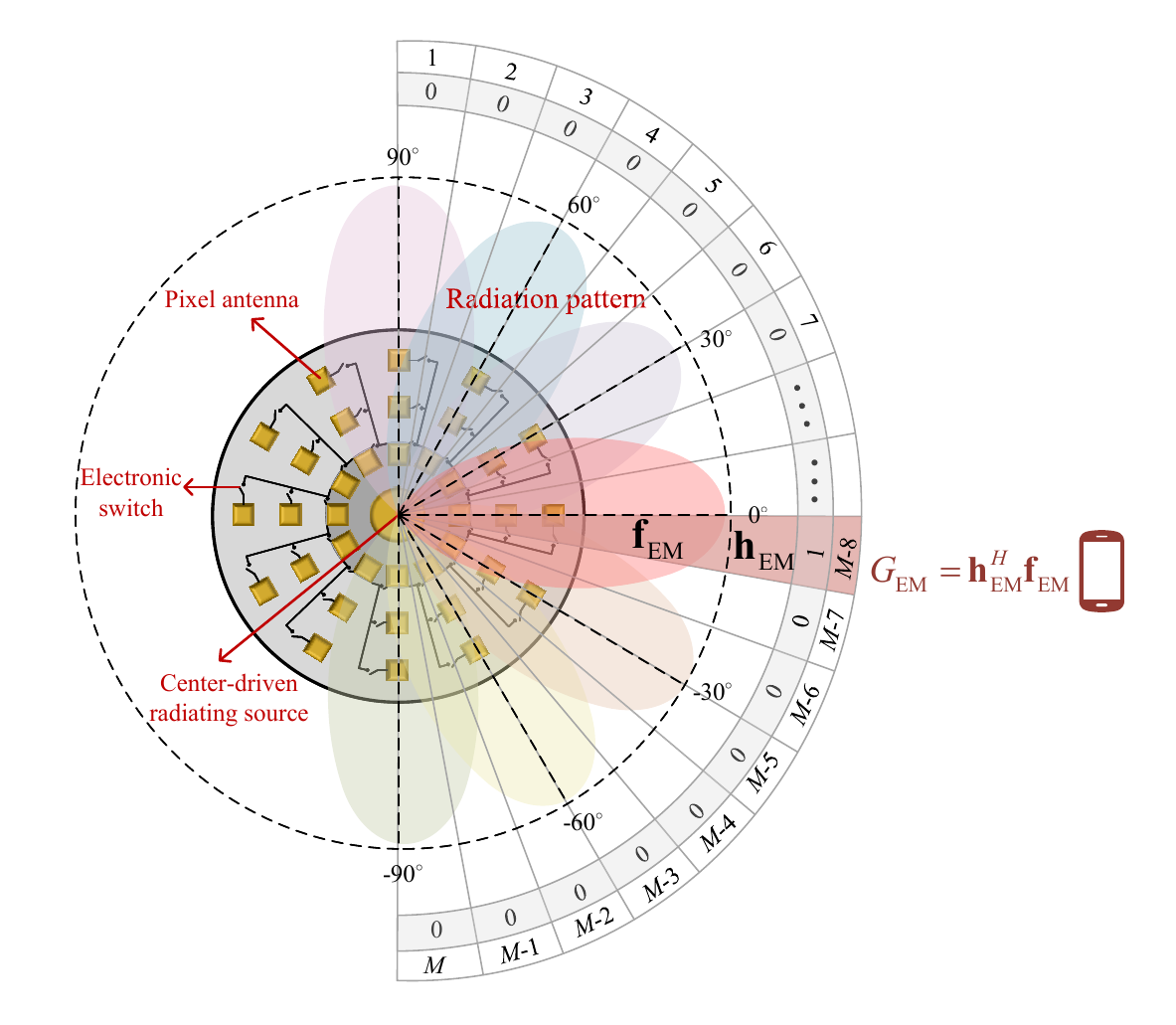}
  \caption{The illustration of a pattern-reconfigurable antenna.}
  \label{fig:reconfigurable_antenna}
   \vspace{-0.4 cm}
\end{figure}

\subsection{Pattern-reconfigurable Antenna}
Pattern-reconfigurable antennas represent advanced technologies that dynamically adapt their radiation patterns, significantly enhancing signal quality, mitigating interference, and improving coverage in wireless communication systems. Several hardware implementations have been explored to achieve such flexible radiation control, notably ESPAR antennas \cite{R. Murch 2023 ESPAR}, FSS \cite{FSS}, and pixel antennas \cite{K.K.Wong 2025 pixel}. Among these, pixel antenna technology has emerged as the most practical and versatile solution, particularly suitable for integration into compact antenna arrays. Hardware prototypes of pixel-based reconfigurable antennas are illustrated in Fig. \ref{fig:hardware}. This prototype employs PIN-diode switching networks that exhibit microsecond-level switching latencies and highly reliable reconfiguration behavior, confirming their suitability for rapid EM domain adaptation within the proposed framework. Moreover, these antennas typically require only a minimal number of serialized binary control lines and low-voltage direct-current (DC) bias circuits, resulting in affordable control signaling overhead and power consumption.

Having established their practicality for real-world deployment, we now detail the hardware configuration and operating principles of pixel-based pattern-reconfigurable antennas. As depicted in Fig. \ref{fig:reconfigurable_antenna}, pixel antenna technology discretizes a continuous antenna surface into numerous electrically small parasitic elements, termed pixels, driven centrally by an active radiating source. These pixels, functioning as fundamental building blocks, employ low-voltage, binary-controlled PIN diodes to connect or disconnect adjacent elements. By adjusting these switches' states, various pixel topologies are formed, modifying the current distribution across the antenna surface. This reconfiguration process effectively alters the radiated electric and magnetic field distributions, thus enabling the generation of diverse  radiation patterns.

To rigorously evaluate the performance improvements brought by pattern-reconfigurable antennas, it is crucial to establish a mathematical model capturing both radiation patterns and corresponding channel characteristics. As aforementioned, pattern-reconfigurable antennas introduce an additional DoF in the EM domain, effectively expanding conventional antennas' spatial representation from a one-dimensional model to an $M$-dimensional angular domain. As illustrated in Fig. \ref{fig:reconfigurable_antenna}, the angular space $[-\pi/2, \pi/2]$ is uniformly divided into $M$ sampling points, which can be expressed as $\vartheta_m =-\frac{\pi}{2}+ \frac{\pi m}{M},~m = 1,2,\ldots, M$.
Specifically, each reconfigurable radiation pattern (illustrated by the red pattern in Fig. \ref{fig:reconfigurable_antenna}) is represented across $M$ discrete angular directions by a vector
\begin{equation}
\mathbf{f}_{\mathrm{EM}}\in \mathbb{R}^{M},
\end{equation}
with each element corresponding to radiation gain in a particular direction. Correspondingly, we introduce the virtual angular index channel for pattern-reconfigurable antennas, denoted as
\begin{equation}
\mathbf{h}_{\mathrm{EM}}\in\{ 0,1 \}^{M},
\end{equation}
which serves as an indexing vector indicating the spatial angular information of the user. Consequently, the effective EM domain gain can be mathematically expressed as
\begin{equation}
G_{\mathrm{EM}}=\mathbf{h}_{\mathrm{EM}}^{H}\mathbf{f}_{\mathrm{EM}}\in \mathbb{R},
\end{equation}
which quantifies the radiation intensity toward the user's actual spatial angle. By optimizing the radiation pattern $\mathbf{f}_{\mathrm{EM}}$ to closely align with the angular channel    $\mathbf{h}_{\mathrm{EM}}$, we can maximize their correlation, thus significantly enhancing the overall radiation efficiency. This theoretical formulation establishes a foundation for the subsequent analysis and optimization.

\begin{figure}[!t]
  \centering
  \includegraphics[width= 3.8 in]{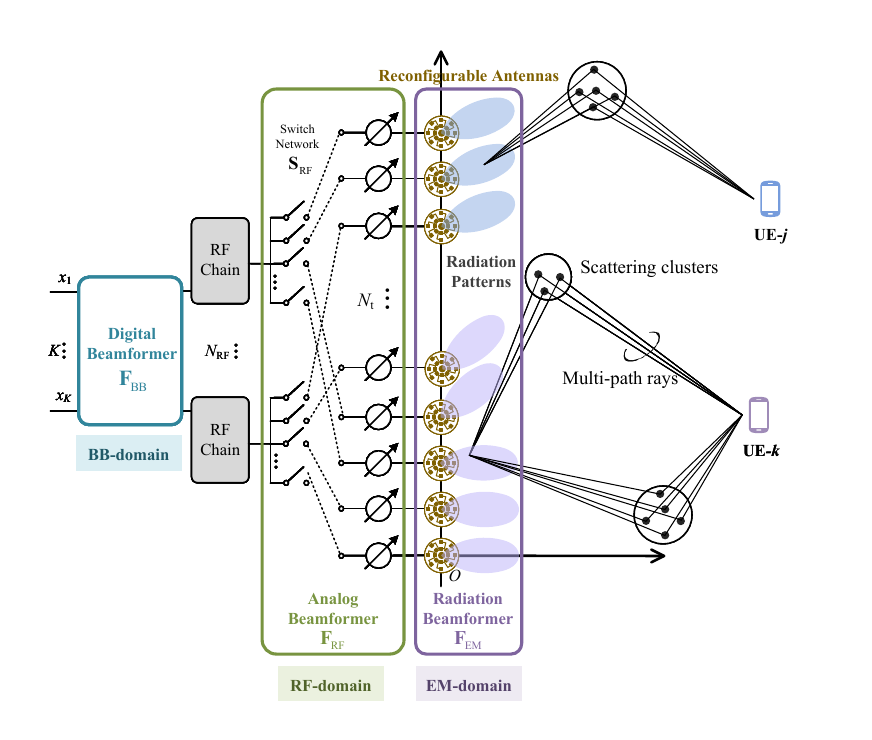}
  \vspace{-0.6cm}
  \caption{Tri-hybrid beamforming architecture.}
  \label{fig:systemmodel}
  \vspace{-0.4cm}
\end{figure}

Inspired by the increased design DoF introduced by pattern-reconfigurable antennas, we propose integrating this advanced antenna technology into conventional hybrid beamforming architectures, and develop a novel tri-hybrid beamforming framework that simultaneously considers three distinct beamforming domains. Specifically, radiation beamforming within the EM domain dynamically shapes the antenna radiation patterns, analog beamforming in the RF domain enhances array gain, and digital beamforming in the BB domain mitigates multi-user interference. This comprehensive integration fully leverages the unique advantages of each domain, thereby achieving superior performance and efficiency in complex wireless communication scenarios.

\vspace{-0.3cm}
\subsection{Tri-hybrid Beamforming Architecture}
To provide a clear illustration of the proposed tri-hybrid beamforming architecture, we consider a mmWave MU-MISO downlink communication system, as shown in Fig. \ref{fig:systemmodel}. Specifically, a base station (BS) equipped with a uniform linear array (ULA) serves $K$ single-antenna user equipments (UEs). The wireless channel between the BS and each UE comprises $L$ propagation path. The BS employs $N_{\mathrm{t}}$ reconfigurable antennas that dynamically adjust their radiation patterns, while each UE is equipped with a conventional omnidirectional antenna. The antenna spacing at the BS is set to $d={\lambda}/{2}$, where $\lambda$ is the center carrier wavelength.
In the following, we mathematically formulate the radiation beamformer $\mathbf{F}_{\mathrm{EM}}$ in the EM domain, the analog beamformer  $\mathbf{F}_{\mathrm{RF}}$ in the RF domain and the digital beamformer  $\mathbf{F}_{\mathrm{BB}}$ in the BB domain, establishing the foundation for the tri-hybrid beamforming architecture.

\textbf{EM domain:}
The radiation beamformer $\mathbf{F}_{\mathrm{EM}}$ dynamically adjusts the amplitude responses of each reconfigurable antenna element, facilitating flexible control over energy distribution across diverse spatial directions. Extending the radiation pattern model $\mathbf{f}_{\mathrm{EM}}$ described in Section II-A for a single pattern-reconfigurable antenna, we characterize the collective programmable radiation pattern of the entire antenna array at the BS, comprising
$N_{\mathrm{t}}$ reconfigurable antennas.
Mathematically, the EM domain radiation beamformer is represented in the following matrix form
\begin{equation}\label{eq:F_EM_extended}   \!\!\mathbf{F}_{\mathrm{EM}}\triangleq\mathrm{blkdiag}\{\mathbf{f}_{\mathrm{EM},1}, \mathbf{f}_{\mathrm{EM},2}, \ldots, \mathbf{f}_{\mathrm{EM},N_{\mathrm{t}}}\}\in\mathbb{R}^{MN_{\mathrm{t}}\times N_{\mathrm{t}}},
\end{equation}
where the $m$-th element of $\mathbf{f}_{\mathrm{EM},n}$ represents the radiation gain of the $n$-th antenna in the $m$-th angular direction. To ensure fairness in pattern design, the total energy of each antenna's radiation pattern is restricted and normalized as $\|\mathbf{f}_{\mathrm{EM},n}\|^{2}=1$.
To further delineate the spatial characteristics of reconfigurable antennas, we introduce the virtual angular index channel matrix associated with the $l$-th propagation path for the $k$-th UE
\begin{equation}\label{eq:H_EM_extended}
\mathbf{H}_{\mathrm{EM},k,l}\triangleq\mathbf{I}_{N_{\mathrm{t}}}\otimes\mathbf{h}_{\mathrm{EM},k,l}\in\{0,1\}^{MN_{\mathrm{t}}\times N_{\mathrm{t}}},~\forall k,l,
\end{equation}
where
$\mathbf{h}_{\mathrm{EM},k,l}\in\{0,1\}^{M}$ explicitly specifies the spatial angles associated with the $k$-th UE. Each $\mathbf{h}_{\mathrm{EM},k,l}$ contains exactly one non-zero element, indicating the angular direction of the $l$-th propagation path for the $k$-th UE among $M$ angular sampling points.
Consequently, the actual radiation gains of all antennas $\mathbf{G}_{\mathrm{EM},k,l} \in\mathbb{R}^{N_{\mathrm{t}}\times N_{\mathrm{t}}} $ can be formulated as
\begin{equation}\label{eq:G}
\begin{aligned}    \mathbf{G}_{\mathrm{EM},k,l}&=\mathbf{H}_{\mathrm{EM},k,l}^{H}\mathbf{F}_{\mathrm{EM}} \\
    &\triangleq \mathrm{diag}\{G_{\mathrm{EM},k,l,1}, \ldots,G_{\mathrm{EM},k,l,N_{\mathrm{t}}}\},~\forall k,l,
\end{aligned}
\end{equation}
which is a diagonal matrix  with the $n$-th entry $G_{\mathrm{EM},k,l,n}\triangleq\mathbf{h}_{\mathrm{EM},k,l}^{H}\mathbf{f}_{\mathrm{EM},n}$ representing the actual radiation gain of the $n$-th antenna towards the $l$-th path of the $k$-th UE. In summary, the matrices $\mathbf{F}_{\mathrm{EM}}$  and $\mathbf{H}_{\mathrm{EM},k,l}$  mathematically capture the additional DoF afforded by reconfigurable antennas, enabling highly adaptive radiation beamforming in the EM domain.

\textbf{RF domain:}
The analog beamformer $\mathbf{F}_{\mathrm{RF}}$ controls the phases of the transmitted signals across antennas, ensuring coherent signal combination and directional beamforming toward desired UEs. We adopt a dynamic-subarray analog beamforming architecture, wherein each RF chain adaptively connects to distinct subsets of antennas through a switching network. This architecture optimally balances system efficiency and communication performance by exploiting dynamic antenna selection. Assuming the number of RF chains $N_{\mathrm{RF}}$ meets or exceeds the number of UEs $K$, i.e., $N_{\mathrm{RF}}\geq K$. The analog beamformer employs a set of low-resolution PSs, each following a constant modulus constraint with quantized phase control of $B$ bits. The discrete phase set of PSs is given by  $\mathcal{F}\triangleq \{\frac{1}{\sqrt{N_{\mathrm{t}}}}e^{\jmath\frac{2\pi b}{2^{B}}}|b=0, 1, \ldots, 2^{B}-1\}$. To facilitate dynamic-subarray beamforming, we define the analog beamformer matrix as $\mathbf{F}_{\mathrm{RF}}\in \{\mathcal{F},0\}^{N_{\mathrm{t}}\times N_{\mathrm{RF}}}$, which implies that if the $l_{\mathrm{RF}}$-th RF chain is connected to the $n$-th antenna, the corresponding phase-shift takes a non-zero value, i.e., $\mathbf{F}_{\mathrm{RF}}(n,l_{\mathrm{RF}})\in \mathcal{F}$; if not, it is set as zero, i.e., $\mathbf{F}_{\mathrm{RF}}(n,l_{\mathrm{RF}})=0$. Moreover, to ensure non-overlapping subarrays, each antenna is dynamically assigned to only one RF chain, meaning that the analog beamformer matrix is restricted to have only one non-zero element per row, i.e., $\|\mathbf{F}_{\mathrm{RF}}(n,:)\|_{0} = 1$.

\textbf{BB domain:}
The digital beamformer $\mathbf{F}_{\mathrm{BB}}$ operates at the baseband level to effectively suppress multi-user interference and optimize overall system performance. Specifically, it is defined as   $\mathbf{F}_{\mathrm{BB}}\triangleq[\mathbf{f}_{\mathrm{BB},1}, \mathbf{f}_{\mathrm{BB},2}, \ldots, \mathbf{f}_{\mathrm{BB},K}]\in \mathbb{C}^{N_{\mathrm{RF}}\times K}$. By performing precise pre-processing of transmitted signals, the digital beamformer compensates for the limited number of RF chains, ensuring refined control over each user's signal in the spatial domain and significantly enhancing system throughput.

\vspace{-0.6cm}
\subsection{Channel Model}
After establishing the mathematical formulation of the tri-hybrid beamforming architecture, our focus now shifts to developing a comprehensive channel model that accommodates the expanded dimensionality introduced by the pattern-reconfigurable antenna array. Specifically, unlike conventional channel models that primarily account for path loss and spatial phase gradients, our proposed model explicitly integrates the virtual angular index channel matrix  $\mathbf{H}_{\mathrm{EM},k,l}$, defined in \eqref{eq:H_EM_extended}, to represent the angular characteristics of each propagation path. By incorporating this channel matrix along with the radiation beamformer $\mathbf{F}_{\mathrm{EM}}$ described in \eqref{eq:F_EM_extended}, our approach precisely determines the radiation gain  $\mathbf{G}_{\mathrm{EM},k,l}$ in \eqref{eq:G}, enabling a more accurate and comprehensive characterization of the EM domain behavior.

To establish a foundation for the innovative channel model of the pattern-reconfigurable antenna array, we first revisit the traditional mmWave channel model. Given the sparse scattering characteristics of the mmWave band, the channel typically consists of $C$ dominant scattering clusters, each containing $L_{c}$ multi-path rays \cite{scattering cluster channel}, \cite{scattering cluster channel2}. The channel between the $n$-th antenna and the $k$-th UE can be expressed as
\begin{equation}\label{eq:channel_cluster}
h_{k,n}=\sum_{c=1}^{C}\sum_{l_{c}=1}^{L_{c}}\alpha_{k,c}e^{j\phi_{k,l_{c}}}e^{-\jmath\frac{2\pi}{\lambda}(n-1)d\sin\theta_{k,c,l_{c}}},~\forall k,n,
\end{equation}
where $\theta_{k,c,l_{c}}$ denotes the angle of departure (AoD) corresponding to the $l_{c}$-th ray in the $c$-th scattering cluster, respectively, $\alpha_{k,c}$ denotes the mean large-scale channel gain of the $c$-th scattering cluster, and $\phi_{k,l_{c}}$ denotes the small-scale phase term of the $l_{c}$-th ray. The AoD $\theta_{k,c,l_{c}}$ can be described as
\begin{equation}\label{eq:theta rays calculation}
   \theta_{k,c,l_{c}} = \overline{\theta}_{k,c} + \Delta\theta_{k,l_{c}}, ~\forall k,c,l_{c},
\end{equation}
where $\overline{\theta}_{k,c}$ denotes the nominal AoD of the $c$-th scattering cluster for the $k$-th UE, and $\Delta\theta_{k,l_{c}}$ denotes the angular deviation of the $l_{c}$-th ray from the nominal direction, following Gaussian distribution $\Delta\theta_{k,l_{c}}\sim\mathcal{N}(0,\varsigma^{2}_{k,c})$, with $\varsigma_{k,c}$ indicating the angle spread. The mean large-scale channel gain $\alpha_{k,c}$ mainly involves the typically distance-dependent path loss with $r_{k,c}$ representing the propagation distance. The small-scale parameter $\phi_{k,l_{c}}$ follows a uniform distribution in $[0, 2\pi]$ and is independent from ray to ray.
For better clarity in the subsequent channel model formulation, we define $L \triangleq\sum_{c=1}^{C}L_{c}$ as the total number of paths. With this notation, the channel model in \eqref{eq:channel_cluster} can be reformulated in a more compact form as
\vspace{-0.2cm}
\begin{equation} h_{k,n}=\sum_{l=1}^{L}\alpha_{k,l}e^{\jmath\phi_{k,l}}
  e^{-\jmath\frac{2\pi}{\lambda}(n-1)d\sin\theta_{k,l}}, ~\forall k,n.
\end{equation}
\vspace{-0.4cm}

Furthermore, the integration of pattern-reconfigurable antennas extends the channel representation from a one-dimensional to an $M$-dimensional formulation by incorporating additional EM domain channel information, i.e., $\mathbf{H}_{\mathrm{EM},k,l}$ in \eqref{eq:H_EM_extended}. Specifically, the dimension-extended channel for the $n$-th pattern-reconfigurable antenna $\mathbf{h}_{k,n} \in \mathbb{C}^{M}$ is expressed as
\begin{equation}
  \!\!\!\mathbf{h}_{k,n}\!\!=\!\!\sum_{l=1}^{L}(\alpha_{k,l}
  e^{\jmath\phi_{k,l}}e^{-\jmath\frac{2\pi}{\lambda}(n-1)d\sin\theta_{k,l}}\!)\!\otimes\!\mathbf{h}_{\mathrm{EM},k,l},~\forall k,n.\!\!\!\!
\end{equation}
Following this formulation, by stacking the channel vectors $\mathbf{h}_{k,n}$ of all antennas, we obtain the comprehensive channel vector for the $k$-th UE, given by $\mathbf{h}_{k}\triangleq[\mathbf{h}_{k,1}^{T}, \mathbf{h}_{k,2}^{T}, \ldots, \mathbf{h}_{k,N_{\mathrm{t}}}^{T}]^{T}\in \mathbb{C}^{MN_{\mathrm{t}}}$. The corresponding expression can be  explicitly formulated as
\begin{equation}\label{eq:expansion_channel_hk}
\begin{aligned}
    \mathbf{h}_{k}&=\sum_{l=1}^{L}(\alpha_{k,l}e^{\jmath\phi_{k,l}}
  \mathbf{a}_{k,l})\otimes \mathbf{h}_{\mathrm{EM},k,l}, \\
                &\triangleq\sum_{l=1}^{L}\underbrace{(\mathbf{I}_{N_{\mathrm{t}}}\otimes\mathbf{h}_{\mathrm{EM},k,l})}_{\mathbf{H}_{\mathrm{EM},k,l}}\underbrace{(\alpha_{k,l}e^{\jmath\phi_{k,l}}\mathbf{a}_{k,l})}_{\mathbf{h}_{\mathrm{S},k,l}}, ~\forall k,\\
\end{aligned}
\end{equation}
where the steering vector is defined as
\begin{equation}
\begin{aligned}
   \!\!\!\mathbf{a}_{k,l}&\!\triangleq\![1, e^{-\jmath\frac{2\pi}{\lambda}d\sin\theta_{k,l}}, \ldots, e^{-\jmath\frac{2\pi}{\lambda}(N_{\mathrm{t}}-1)d\sin\theta_{k,l}}]^{T},~\forall k,l.\!\!\!\!\!\!\!\!
\end{aligned}
\end{equation}
The comprehensive channel $\mathbf{h}_{k}$ in \eqref{eq:expansion_channel_hk} is composed of the virtual angular index channel matrix $\mathbf{H}_{\mathrm{EM},k,l}$ defined in \eqref{eq:H_EM_extended} and the spatial domain channel $\mathbf{h}_{\mathrm{S},k,l}$ given by
\begin{equation}
\mathbf{h}_{\mathrm{S},k,l}\triangleq\alpha_{k,l}e^{\jmath\phi_{k,l}}\mathbf{a}_{k,l},~\forall k,l.
\end{equation}
When the virtual angular index channel matrix $\mathbf{H}_{\mathrm{EM},k,l}$ interacts with the radiation beamformer $\mathbf{F}_{\mathrm{EM}}$ in \eqref{eq:F_EM_extended}, the actual radiation gains of the antennas $\mathbf{G}_{\mathrm{EM},k,l}$ mentioned in \eqref{eq:G} are obtained. These gains subsequently modify the conventional spatial domain channel $\mathbf{h}_{\mathrm{S},k,l}$, thereby shaping the effective channel seen by the analog and digital beamformers.
Additionally, to enhance clarity and brevity, we replace the path-wise summation in \eqref{eq:expansion_channel_hk} with an equivalent block-stacked multiplicative expressions, which yields
\begin{equation}\label{eq:H_EM_estimiate_Cluster}
\mathbf{h}_{k} \triangleq \mathbf{H}_{\mathrm{EM},k}\mathbf{h}_{\mathrm{S},k}, ~\forall k,
\end{equation}
where $\mathbf{H}_{\mathrm{EM},k}\triangleq[\mathbf{H}_{\mathrm{EM},k,1},\mathbf{H}_{\mathrm{EM},k,2}, \ldots,\mathbf{H}_{\mathrm{EM},k,L}]\in\{0,1\}^{MN_{\mathrm{t}}\times LN_{\mathrm{t}}}$ contains the virtual angular index channel matrices, and $\mathbf{h}_{\mathrm{S},k}\triangleq[\mathbf{h}_{\mathrm{S},k,1}^{T},\mathbf{h}_{\mathrm{S},k,2}^{T}, \ldots, \mathbf{h}_{\mathrm{S},k,L}^{T}]^{T}\in\mathbb{C}^{LN_{\mathrm{t}}}$ includes the spatial domain channel information.

\vspace{-0.3cm}
\subsection{System Model}
In the MU-MISO downlink system shown in Fig. \ref{fig:systemmodel}, the transmitted signal is first processed by the digital beamformer $\mathbf{F}_{\mathrm{BB}}$ in the BB domain, followed by the analog beamformer $\mathbf{F}_{\mathrm{RF}}$ in the RF domain and subsequently shaped by the radiation beamformer $\mathbf{F}_{\mathrm{EM}}$ in the EM domain, before being radiated by the pattern-reconfigurable antennas. After propagating through the spatial channel, it is finally received by the UE. Based on the above mathematical definition, the signal received by the $k$-th UE is given by
\begin{equation}\label{eq:received signal}
\begin{aligned}
y_{k}&=\mathbf{h}_{k}^{H}\mathbf{F}_{\mathrm{EM}}\mathbf{F}_{\mathrm{RF}}\mathbf{F}_{\mathrm{BB}}\mathbf{x}+n_{k}\\
&=\mathbf{h}_{\mathrm{S},k}^{H}\mathbf{H}_{\mathrm{EM},k}^{H}\mathbf{F}_{\mathrm{EM}}\mathbf{F}_{\mathrm{RF}}\mathbf{F}_{\mathrm{BB}}\mathbf{x}+n_{k},~\forall k,
\end{aligned}
\end{equation}
where $\mathbf{x}\triangleq[x_{1}, x_2, \ldots, x_{K}]^T\in \mathbb{C}^{K}$ denotes the transmitted symbols of $K$ UEs that satisfies $\mathbb{E}\{\mathbf{x}\mathbf{x}^{H}\}=\mathbf{I}_{K}$, and $n_{k}\!\sim\!\mathcal{CN}(0,\sigma^{2}_{k})$ denotes the additive white Gaussian noise (AWGN) at the $k$-th UE.
The signal-to-interference-plus-noise ratio (SINR) of the $k$-th UE can be calculated as
\begin{equation}\label{eq:SINR}
\mathrm{SINR}_{k}=\frac{|\mathbf{h}_{\mathrm{S},k}^{H}\mathbf{H}_{\mathrm{EM},k}^{H}\mathbf{F}_{\mathrm{EM}}\mathbf{F}_{\mathrm{RF}}\mathbf{f}_{\mathrm{BB},k}|^{2}}{\sum_{j=1,j\neq k}^{K}|\mathbf{h}_{\mathrm{S},k}^{H}\mathbf{H}_{\mathrm{EM},k}^{H}\mathbf{F}_{\mathrm{EM}}\mathbf{F}_{\mathrm{RF}}\mathbf{f}_{\mathrm{BB},j}|^{2}+\sigma_{k}^{2}}, ~\forall k,\\
\end{equation}
and the corresponding achievable rate can be calculated as
\begin{equation}\label{eq:sum-rate}
R_{k}=\log_{2}(1+\mathrm{SINR}_{k}).
\end{equation}

\vspace{-0.9cm}
\subsection{Problem Formulation}
In this paper, we propose a comprehensive tri-hybrid beamforming strategy, jointly optimizing the radiation beamformer $\mathbf{F}_{\mathrm{EM}}$ in the EM domain, the analog beamformer $\mathbf{F}_{\mathrm{RF}}$ in the RF domain, and the digital beamformer $\mathbf{F}_{\mathrm{BB}}$ in the BB domain. The primary objective is to maximize the sum-rate performance of the system while satisfying several critical constraints, including radiation pattern energy normalization, constant modulus and discrete-phase constraints of the PSs, non-overlapping dynamic subarray assignments, and a total transmit power budget. Therefore, the joint optimization problem encompassing the three beamforming domains under the assumption of perfect instantaneous CSI is formulated as
\begin{subequations}
\label{eq:original_problem}
\begin{align}
\label{eq:original_problem_a}\max_{\mathbf{F}_{\mathrm{EM}},\mathbf{F}_{\mathrm{RF}}, \mathbf{F}_{\mathrm{BB}}}&\ \ \sum^{K}_{k=1}R_{k}\\
     \mathrm{s.t.}\ \ \ \ \
     \label{eq:original_problem_b}
    &\ \ \|\mathbf{f}_{\mathrm{EM},n}\|^{2}=1,~\forall n,\\
    \label{eq:original_problem_c}
    &\ \ \mathbf{F}_{\mathrm{RF}}(n,l_{\mathrm{RF}})\in \{\mathcal{F},0\}, ~\forall n,l_{\mathrm{RF}},\\
    \label{eq:original_problem_d}
    &\ \ \|\mathbf{F}_{\mathrm{RF}}(n,:)\|_{0}=1, ~\forall n,\\
    \label{eq:original_problem_e}
    &\ \ \|\mathbf{F}_{\mathrm{RF}}\mathbf{F}_{\mathrm{BB}}\|_{F}^{2}\leq P_{\mathrm{t}},
\end{align}
\end{subequations}
where $P_{\mathrm{t}}$ is the transmit power budget.


Accurately obtaining instantaneous CSI poses significant practical difficulties within this tri-hybrid framework. In the RF domain, the analog beamformer inherently suffers from undersampling due to the limited number of RF chains, while acquiring instantaneous CSI for all possible radiation patterns in the EM domain leads to excessive pilot overhead. In addition, real-time joint optimization of the beamforming matrices across all three domains results in high computational complexity and low processing efficiency. These challenges motivate the development of a more structured beamforming strategy, which will be addressed in the next section.

\begin{figure*}[!t]
  \centering
  \includegraphics[width= \linewidth]{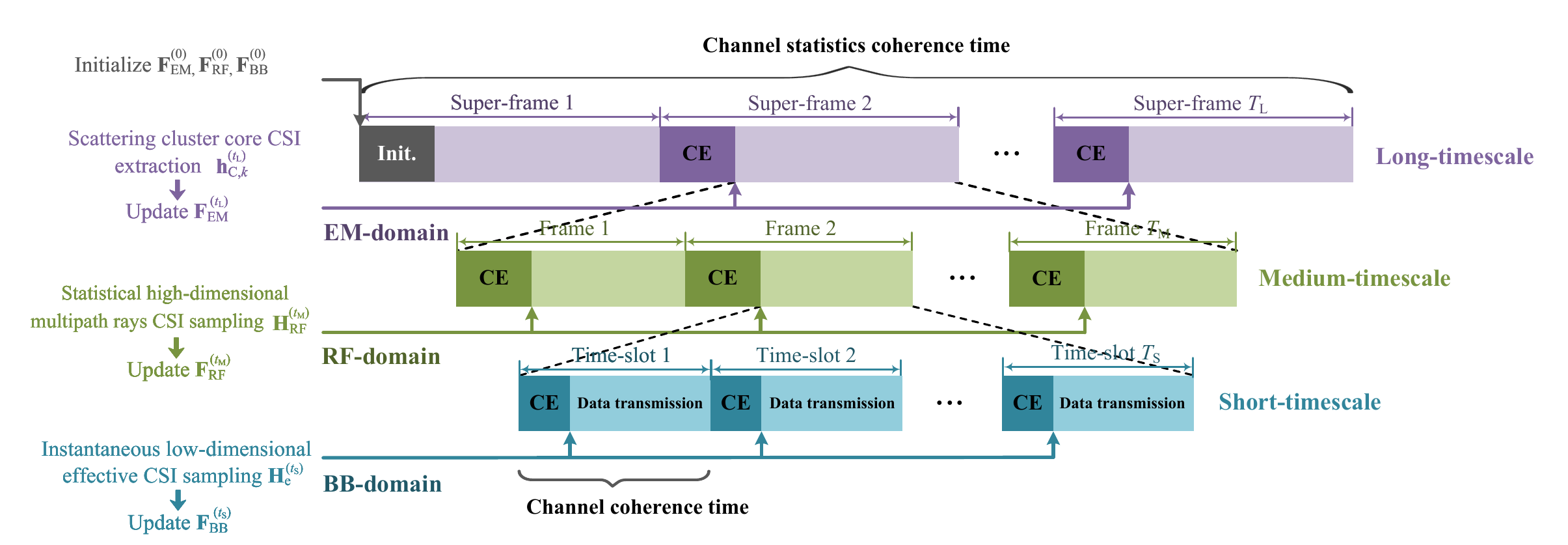}
  \caption{ The frame structure of tri-timescale tri-hybrid beamforming.}
  \label{fig:S-CSI}
\end{figure*}

\section{Tri-timescale Beamforming Designs}
\subsection{Tri-timescale Framework}
In this section, we present a tri-timescale beamforming framework for the proposed tri-hybrid architecture. Instead of estimating full instantaneous CSI and jointly optimizing the radiation, analog, and digital beamformers in every time-slot, the design separates their updates across three different timescales so as to balance performance and implementation complexity, as illustrated in Fig.~\ref{fig:S-CSI}.
The overall time horizon corresponds to the channel statistics coherence time, during which the statistical distribution of the channel remains approximately constant\footnote{Our channel modeling follows the quasi-static clustered model described in 3GPP TR 38.901 \cite{scattering cluster channel}, where cluster-level parameters (e.g., nominal angles and angle spreads) remain approximately constant within a channel-statistics coherence interval. This assumption is well suited for low-to-moderate mobility scenarios such as indoor hotspots and small-cell mmWave deployments. Extending the proposed tri-timescale framework to fully dynamic, high-mobility scenarios is an important topic for future research.}. This period is organized hierarchically, with beamforming operations carried out over three distinct timescales, as detailed below:
\begin{itemize}
  \item \textbf{Long-timescale}: The total duration, corresponding to the channel-statistics coherence time, is divided into $T_{\mathrm{L}}$ super-frames, denoted as $\mathcal{T}_{\mathrm{L}}\triangleq\{1,2,\ldots,T_{\mathrm{L}}\}$. This coherence interval typically spans from a few seconds to several tens of seconds.  In each super-frame, the BS extracts high-dimensional channel information of the scattering cluster core to design the radiation beamformer, ensuring radiation pattern alignment in the EM domain.
  \item \textbf{Medium-timescale}: Each super-frame consists of $T_{\mathrm{M}}$ frames, denoted as $\mathcal{T}_{\mathrm{M}}\triangleq\{1,2,\ldots,T_{\mathrm{M}}\}$. The duration of each frame is determined by the sampling frequency of the statistical channel information, generally ranging from several hundred milliseconds to a few seconds.  In each frame, the BS acquires a statistical channel sample of multi-path rays within the scattering clusters to update the analog beamformer, enhancing array gain in the RF domain.
  \item \textbf{Short-timescale}: Each frame is further divided into $T_{\mathrm{S}}$ time-slots, denoted as $\mathcal{T}_{\mathrm{S}}\triangleq\{1,2,\ldots,T_{\mathrm{S}}\}$. Each time-slot is referred to as the channel coherence time, typically on the order of tens of milliseconds\footnote{The durations of the three timescales depend on the carrier frequency, user mobility, and propagation environment. The orders of magnitude reported here follow typical mmWave deployments under the 3GPP TR 38.901 clustered channel model and serve only as representative examples [29]. In practice, the time-scale durations can be flexibly adjusted to match specific deployment scenarios without altering the proposed tri-timescale optimization framework.}. In each time-slot, the BS obtains an instantaneous low-dimensional effective CSI to refine the digital beamformer, mitigating multi-user interference and accounting for fast channel fluctuations in the BB domain.
\end{itemize}

This hierarchical framework significantly reduces channel estimation overhead while ensuring efficient beamforming adaptation across different domains and timescales. In the following subsections, we provide a detailed description of the channel estimation processes and beamforming design procedures based on this tri-timescale framework.

\subsection{Long-timescale EM Domain Design}
\subsubsection{Long-timescale EM domain Channel Estimation}
Given the specific scattering cluster characteristics described in \eqref{eq:channel_cluster}, the angles of the multi-path rays within each scattering cluster exhibit only limited fluctuations around the cluster core. This observation implies that real-time radiation patterns adaptation based on instantaneous CSI is unnecessary. Instead, aligning radiation patterns with the nominal angles of scattering clusters is sufficient to maintain robust communication performance, while significantly reducing channel estimation overhead and beamforming design complexity.
Therefore, over the long-timescale, our objective is to estimate the nominal angles $\overline{\theta}_{k,c}$ and the mean large-scale channel gains $\alpha_{k,c}$ of the scattering clusters, which serve as the foundation for radiation beamforming optimization.

Specifically, we employ the Capon algorithm to construct the angular power spectrum (APS) based on the channel covariance matrix, enabling the extraction of scattering cluster channel parameters \cite{scattering cluster channel2}, \cite{F. Gao 2016 Scattering cluster}. To illustrate the channel estimation process, we consider the $k$-th UE as an example. In the first super-frame, all the pattern-reconfigurable antennas are typically configured to radiate toward the boresight direction. Starting from the second super-frame, the scattering-cluster channel information is estimated by accumulating spatial channel samples collected over the medium-timescale of the preceding super-frames. Accordingly, the channel covariance matrix at the $t_{\mathrm{L}}$-th super-frame is given by
\begin{equation}\label{eq:channel correlated matrix}
\mathbf{R}_{k}^{(t_{\mathrm{L}})}=\frac{1}{t_{\mathrm{L}}T_{\mathrm{M}}}\sum_{t_{\mathrm{l}}=1}^{t_{\mathrm{L}}-1}\sum_{t_{\mathrm{M}}=1}^{T_{\mathrm{M}}}
\widetilde{\mathbf{h}}^{(t)}_{\mathrm{S},k}(\widetilde{\mathbf{h}}_{\mathrm{S},k}^{(t)})^{H},~\forall k,t_{\mathrm{L}},
\end{equation}
where $t=(t_{\mathrm{l}}-1)T_{\mathrm{M}}+t_{\mathrm{M}}$ is defined for brevity and $\widetilde{\mathbf{h}}_{\mathrm{S},k}\triangleq\sum_{l}^{L}\alpha_{k,l}e^{j\phi_{k,l}}\mathbf{a}_{k,l}$ represents the estimated channel excluding the influence of the radiation patterns.
Then, the APS can be derived as
\begin{equation}\label{eq:APS}
\rho^{(t_{\mathrm{L}})}_{k}(\vartheta_{m})=\frac{1}{\mathbf{e}^{H}(\vartheta_{m})(\mathbf{R}_{k}^{(t_{\mathrm{L}})})^{-1}\mathbf{e}(\vartheta_{m})},~\forall k,t_{\mathrm{L}},
\end{equation}
with $
   \mathbf{e}(\vartheta_{m})\!\triangleq\![1, e^{-\jmath\frac{2\pi}{\lambda}d\sin\vartheta_{m}}, \ldots, e^{-\jmath\frac{2\pi}{\lambda}(N_{\mathrm{t}}-1)d\sin\vartheta_{m}}]^{T}\in\mathbb{C}^{N_{\mathrm{t}}}.
$
Since the APS reflects the power distribution across the angular domain, each peak in the APS physically represents the presence of a distinct scattering cluster in the propagation environment. Thus, the number of the peaks in $\rho^{(t_{\mathrm{L}})}_{k}(\vartheta_{m})$ that exceed a specified threshold directly provides the estimated number of scattering clusters $C$. For each of these peaks, the corresponding angle value and power value indicate the estimated nominal angle $\overline{\theta}_{k,c}$ and the mean large-scale channel gain $\alpha_{k,c}$ of the corresponding scattering cluster, respectively. In conclusion, referring to \eqref{eq:H_EM_estimiate_Cluster}, we can construct the channels of scattering clusters as
\begin{equation}\label{eq:EM_channel}
\mathbf{h}_{\mathrm{C},k}=\mathbf{H}_{\mathrm{EM},\mathrm{C},k}\mathbf{h}_{\mathrm{S},\mathrm{C},k}\in\mathbb{C}^{MN_{\mathrm{t}}},~\forall k,
\end{equation}
where $\mathbf{H}_{\mathrm{EM},\mathrm{C},k}\triangleq[\mathbf{H}_{\mathrm{EM},\mathrm{C},k,1},\mathbf{H}_{\mathrm{EM},\mathrm{C},k,2}, \ldots,\mathbf{H}_{\mathrm{EM},\mathrm{C},k,C}]\in\{0,1\}^{MN_{\mathrm{t}}\times CN_{\mathrm{t}}}$ indicates the EM domain angle information and $\mathbf{h}_{\mathrm{S},\mathrm{C},k}\triangleq[\mathbf{h}_{\mathrm{S},\mathrm{C},k,1}^{T},\mathbf{h}_{\mathrm{S},\mathrm{C},k,2}^{T}, \ldots, \mathbf{h}_{\mathrm{S},\mathrm{C},k,C}^{T}]^{T}\in\mathbb{C}^{CN_{\mathrm{t}}}$ encapsulates the spatial domain channel characteristics of $C$ scattering cluster cores. It serve as the foundation for the long-timescale EM domain radiation beamformer design.

\subsubsection{Long-timescale Radiation Beamformer Design}
Notably, over the long-timescale, the EM domain radiation beamformer $\mathbf{F}_{\mathrm{EM}}$ mainly aims to establish a basic pattern alignment with the scattering cluster core. This optimization requires only the channel information of the scattering cluster core, i.e., the estimated nominal angles and average channel gains of the scattering clusters. Consequently, we substitute the instantaneous multi-path channel $\mathbf{h}_{k}$ in the SINR expression \eqref{eq:SINR} and the sum-rate formulation \eqref{eq:sum-rate} with the scattering cluster core channel $\mathbf{h}_{\mathrm{C},k}$, and define the long-timescale sum-rate $R_{\mathrm{C},k}$ as our optimization objective. Given a fixed analog beamformer $\mathbf{F}_{\mathrm{RF}}$ and digital beamformer $\mathbf{F}_{\mathrm{BB}}$, the long-timescale EM domain radiation beamforming design problem is then formulated as
\begin{subequations}
\label{eq:original_problem_EM}
\vspace{-0.2cm}
\begin{align}
\label{eq:original_problem_EM_a}
    \max_{\mathbf{F}_{\mathrm{EM}}}&\ \ \sum^{K}_{k=1}R_{\mathrm{C},k}\\
     \mathrm{s.t.}
     \label{eq:original_problem_EM_b}
    &\ \ \|\mathbf{f}_{\mathrm{EM},n}\|^{2}=1,~\forall n.
\end{align}
\end{subequations}
\vspace{-0.5cm}

Considering the hardware implementability of pixel-based pattern-reconfigurable antennas, we adopt a practical approach by selecting the optimal radiation pattern from a predefined set, rather than designing arbitrary patterns with unrestricted directions and shapes. This approach ensures a balance between implementation efficiency and real-world hardware constraints.
To be specific, the EM domain radiation beamformer is formulated as
\begin{equation}\label{eq:FEM_decomposition}
\mathbf{F}_{\mathrm{EM}}\triangleq\overline{\mathbf{F}}_{\mathrm{pat}}\mathbf{S}_{\mathrm{EM}},
\end{equation}
where $\overline{\mathbf{F}}_{\mathrm{pat}} \triangleq \mathbf{I}_{N_{\mathrm{t}}}\otimes\mathbf{F}_{\mathrm{pat}}\in\mathbb{R}^{MN_{\mathrm{t}}\times PN_{\mathrm{t}}}$ with $\mathbf{F}_{\mathrm{pat}}\in\mathbb{R}^{M\times P}$ representing a predefined radiation pattern dictionary that contains the radiation gains of $P$ candidate patterns over $M$ spatial sampling angles. The matrix $\mathbf{S}_{\mathrm{EM}} \triangleq\mathrm{blkdiag}\{\mathbf{s}_{\mathrm{EM},1},\mathbf{s}_{\mathrm{EM},2},\ldots,\mathbf{s}_{\mathrm{EM},N_{\mathrm{t}}}\}\in\{0,1\}^{PN_{\mathrm{t}}\times N_{\mathrm{t}}}$ denotes the EM domain pattern selection matrix, where each sub-vector $\mathbf{s}_{\mathrm{EM},n}\in\{0,1\}^{P}$ selects a single optimal pattern from the dictionary for the $n$-th antenna, i.e., $\|\mathbf{s}_{\mathrm{EM},n}\|_{1}=1$.
Based on this formulation, the original problem in \eqref{eq:original_problem_EM} is equivalently transformed into a more practical optimization problem focused on optimizing the EM domain pattern selection matrix $\mathbf{S}_{\mathrm{EM}}$, given by
\begin{subequations}
\label{eq:original_problem_SEM}
\vspace{-0.3cm}
\begin{align}
\label{eq:original_problem_SEM_a}
    \max_{\mathbf{S}_{\mathrm{EM}}}&\ \ \sum^{K}_{k=1}R_{\mathrm{C},k}\\
     \mathrm{s.t.}
     \label{eq:original_problem_SEM_b}
    &\ \ \mathbf{s}_{\mathrm{EM},n}(p)\in \{0,1\},~\forall n,p,\\
    \label{eq:original_problem_SEM_c}
    &\ \ \|\mathbf{s}_{\mathrm{EM},n}\|_{1}=1,~\forall n.
\end{align}
\end{subequations}
 \vspace{-0.5cm}

The non-convex optimization problem in \eqref{eq:original_problem_SEM} poses significant challenges due to the fractional and logarithmic terms in the objective function \eqref{eq:original_problem_SEM_a}, as well as the non-smooth, non-convex Boolean constraint in \eqref{eq:original_problem_SEM_b}. To address these difficulties, we first apply the fractional programming (FP) method to reformulate the objective function \eqref{eq:original_problem_SEM_a} into a polynomial expression, making it more tractable. Subsequently, we transform the Boolean constraint \eqref{eq:original_problem_SEM_b} into a quadratic penalty term with a box constraint and leverage the majorization-minimization (MM) method to facilitate efficient optimization.

Firstly, by employing the Lagrangian dual reformulation and introducing auxiliary variable $\bm{\mu}\triangleq[\mu_{1}, \mu_{2}, \ldots, \mu_{K}]^{T}$, the objective function \eqref{eq:original_problem_SEM} can be transformed to
 \vspace{-0.2cm}
\begin{equation} \label{eq:EM_lagrangiandual}
\begin{aligned}
&\sum_{k=1}^{K}\log_{2}(1+\mu_{k})-\sum_{k=1}^{K}\mu_{k}\\
&\ \ \ \ \ +\sum_{k=1}^{K}\frac{(1+\mu_{k})
|\mathbf{h}_{\mathrm{C},k}^{H}\mathbf{F}_{\mathrm{EM}}\mathbf{F}_{\mathrm{RF}}\mathbf{f}_{\mathrm{BB},k}|^{2}}
{\sum_{j=1}^{K}|\mathbf{h}_{\mathrm{C},k}^{H}\mathbf{F}_{\mathrm{EM}}\mathbf{F}_{\mathrm{RF}}\mathbf{f}_{\mathrm{BB},j}|^{2}+\sigma_{k}^{2}},
\end{aligned}
\end{equation}
which is equivalent to the original objective function \eqref{eq:original_problem_SEM_a} when the auxiliary variable $\bm{\mu}$ has the optimal value
\begin{equation}
\label{eq:mu}
\mu_{k}=\frac{|\mathbf{h}_{\mathrm{C},k}^{H}\mathbf{F}_{\mathrm{EM}}\mathbf{F}_{\mathrm{RF}}\mathbf{f}_{\mathrm{BB},k}|^{2}}
{\sum^{K}_{j=1,j\neq k}|\mathbf{h}_{\mathrm{C},k}^{H}\mathbf{F}_{\mathrm{EM}}\mathbf{F}_{\mathrm{RF}}\mathbf{f}_{\mathrm{BB},j}|^{2}+\sigma_{k}^{2}},~ \forall k.
\end{equation}
However, the summation of fractional terms in \eqref{eq:EM_lagrangiandual} still hinders a straightforward solution. Thus, we further adopt the quadratic transform to convert it into
\begin{equation} \label{eq:EM_quadratic transform}
2\sqrt{1+\mu_{k}}\Re\{\xi_{k}^{\ast}\mathbf{h}_{\mathrm{C},k}^{H}\mathbf{F}_{\mathrm{EM}}\mathbf{F}_{\mathrm{RF}}\mathbf{f}_{\mathrm{BB},k}\}-|\xi_{k}|^{2}D_{k}, ~\forall k,
\end{equation}
where $\bm{\xi} \triangleq [\xi_{1},\xi_{2}, \ldots, \xi_{K}]^{T}$ is the auxiliary variable and $D_{k}\triangleq \sum_{j=1}^{K}|\mathbf{h}_{\mathrm{C},k}^{H}\mathbf{F}_{\mathrm{EM}}\mathbf{F}_{\mathrm{RF}}\mathbf{f}_{\mathrm{BB},j}|^{2}+\sigma_{k}^{2}$ is utilized to simplify the expression. The expression \eqref{eq:EM_quadratic transform} is equivalent to the last fractional term in \eqref{eq:EM_lagrangiandual} when the auxiliary variable $\bm{\xi}$ has the optimal value
\begin{equation}
\label{eq:xi}
\xi_{k}=\frac{\sqrt{1+\mu_{k}}\mathbf{h}_{\mathrm{C},k}^{H}\mathbf{F}_{\mathrm{EM}}\mathbf{F}_{\mathrm{RF}}\mathbf{f}_{\mathrm{BB},k}}{D_{k}}, ~\forall k.
\end{equation}
Based on the above derivation, the objective function can be reformulated as
\begin{equation} \label{eq:EM_reformulated objective}
\begin{aligned}
 &{\sum_{k=1}^{K}}\big[\log_{2}(1+\mu_{k})-\mu_{k}\\
\!\!\! &\ +2\sqrt{1+\mu_{k}}\Re\{\xi_{k}^{\ast}
\mathbf{h}_{\mathrm{C},k}^{H}\mathbf{F}_{\mathrm{EM}}\mathbf{F}_{\mathrm{RF}}\mathbf{f}_{\mathrm{BB},k}\}-|\xi_{k}|^{2}D_{k}\big].
\end{aligned}
\end{equation}
In order to facilitate the subsequent transformation, \eqref{eq:EM_reformulated objective} can be rewritten as the following concise form
\begin{equation} \label{eq:EM_objective_concise form}
\sum_{k=1}^{K}\big[\log_{2}(1+\mu_{k})-\mu_{k}-|\xi_{k}|^{2}\sigma_{k}^{2}\big]+\delta,
\end{equation}
where we define
\begin{equation} \label{eq:delta}
\begin{aligned}
\delta\triangleq & \sum_{k=1}^{K}\big[2\sqrt{1+\mu_{k}}\Re\{\xi_{k}^{\ast}
\mathbf{h}_{\mathrm{C},k}^{H}\mathbf{F}_{\mathrm{EM}}\mathbf{F}_{\mathrm{RF}}\mathbf{f}_{\mathrm{BB},k}\}\\
& \hspace{1 cm}-|\xi_{k}|^{2}\sum_{j=1}^{K}|\mathbf{h}_{\mathrm{C},k}^{H}\mathbf{F}_{\mathrm{EM}}\mathbf{F}_{\mathrm{RF}}\mathbf{f}_{\mathrm{BB},j}|^{2}\big].
\end{aligned}
\end{equation}
Having the optimal auxiliary variables $\bm{\mu}$ and $\bm{\xi}$, maximizing \eqref{eq:EM_objective_concise form} is equivalent to maximizing the term $\delta$.
Consequently, we can recast the optimization problem as
\begin{subequations}
\label{eq:problem_SEM_delta}
\begin{align}
\label{eq:problem_SEM_delta_a}
    \max_{\mathbf{S}_{\mathrm{EM}}}&\ \ \ \delta\\
     \mathrm{s.t.}
     \label{eq:problem_SEM_delta_b}
    &\ \ \ \mathbf{s}_{\mathrm{EM},n}(p)\in \{0,1\},~\forall n,p,\\
    \label{eq:problem_SEM_delta_c}
    &\ \ \ \|\mathbf{s}_{\mathrm{EM},n}\|_{1}=1,~\forall n.
\end{align}
\end{subequations}

After dealing with the objective function, we turn to tackle the neither smooth nor convex Boolean constraint \eqref{eq:problem_SEM_delta_b}.
As stated in \cite{Boolean}, the Boolean constraint \eqref{eq:problem_SEM_delta_b} can be transformed into an optimization problem of maximizing a quadratic term with a box constraint, which yields
\begin{subequations}
\label{eq:Boolean}
\begin{align}
    \label{eq:Boolean_a}
    \max_{\mathbf{S}_{\mathrm{EM}}}&\ \ \mathbf{s}_{\mathrm{EM}}^{T}(\mathbf{s}_{\mathrm{EM}}-\mathbf{1})\\
    \mathrm{s.t. }\label{eq:Boolean_b}
     &\ \ \mathbf{s}_{\mathrm{EM},n}(p)\in[0,1], ~\forall n,p,
\end{align}
\end{subequations}
where $\mathbf{s}_{\mathrm{EM}}\triangleq[\mathbf{s}_{\mathrm{EM},1}^{T},\mathbf{s}_{\mathrm{EM},2}^{T},\ldots,\mathbf{s}_{\mathrm{EM},N_{\mathrm{t}}}^{T}]^{T}\in\mathbb{R}^{PN_{\mathrm{t}}}$ is introduced for brevity and the binary variable $\mathbf{S}_{\mathrm{EM}}$ has been relaxed to be continuous.
Then, we can incorporate the objective function \eqref{eq:Boolean_a} as a penalty term into the problem \eqref{eq:problem_SEM_delta} with a box constraint \eqref{eq:Boolean_b}, which is given by
\begin{subequations}
\label{eq:Boolean_penalty}
\begin{align}
   \label{eq:Boolean_penalty_a}
  \max_{\mathbf{S}_{\mathrm{EM}}}&\ \ \ \delta + \varrho_{1}\mathbf{s}_{\mathrm{EM}}^{T}(\mathbf{s}_{\mathrm{EM}}-\mathbf{1}) \\
    \label{eq:Boolean_penalty_b}
     \mathrm{s.t.}
    &\ \ \ \mathbf{s}_{\mathrm{EM},n}(p)\in [0,1],~\forall n,p,\\
    &\ \ \ \|\mathbf{s}_{\mathrm{EM},n}\|_{1}=1,~\forall n,
\end{align}
\end{subequations}
where $\varrho_{1}$ is the penalty parameter. The newly constructed objective function \eqref{eq:Boolean_penalty_a} contains a non-concave quadratic term related to the variable $\mathbf{S}_{\mathrm{EM}}$. Hence, we employ the MM approach in \cite{MM} to convert it into a linear form at the current point $\mathbf{s}_{\mathrm{EM}}^{(q)}$ in the $q$-th iteration, which can be expressed as
\begin{equation}
  \label{eq:Boolean_MM_EM}
  \mathbf{s}_{\mathrm{EM}}^{T}(\mathbf{s}_{\mathrm{EM}}-\mathbf{1})\geq 2(\mathbf{s}_{\mathrm{EM}}^{(q)})^{T}\mathbf{s}_{\mathrm{EM}}-(\mathbf{s}_{\mathrm{EM}}^{(q)})^{T}\mathbf{s}_{\mathrm{EM}}^{(q)} - \mathbf{s}_{\mathrm{EM}}^{T}\mathbf{1}.
\end{equation}
Finally, the optimization problem for designing the EM domain pattern selection matrix is formulated as
\begin{subequations}
\label{eq:EM problem final}
\begin{align}
    \max_{\mathbf{S}_{\mathrm{EM}}}&\ \ \ \delta + \varrho_{1}(2\mathbf{s}_{\mathrm{EM}}^{(q)}-\mathbf{1})^{T}\mathbf{s}_{\mathrm{EM}} \\
    \label{eq:EM problem final b}
     \mathrm{s.t.}
    &\ \ \ \mathbf{s}_{\mathrm{EM},n}(p)\in [0,1],~\forall n,p,\\
    &\ \ \ \|\mathbf{s}_{\mathrm{EM},n}\|_{1}=1,~\forall n,
\end{align}
\end{subequations}
which is convex and can be easily tackled by off-the-shelf solver CVX. With the optimal EM domain pattern selection matrix $\mathbf{S}_{\mathrm{EM}}$ and the known radiation pattern dictionary $\overline{\mathbf{F}}_{\mathrm{pat}}$, the radiation beamformer is constructed as $\mathbf{F}_{\mathrm{EM}}\triangleq\overline{\mathbf{F}}_{\mathrm{pat}}\mathbf{S}_{\mathrm{EM}}$ in \eqref{eq:FEM_decomposition}.

\subsection{Medium-timescale RF Domain Design}
\subsubsection{Medium-timescale RF Domain Channel Estimation}
The presence of reconfigurable antennas inherently couples the spatial-domain channel with the radiation pattern gains, as characterized in \eqref{eq:H_EM_estimiate_Cluster}. Conventional channel estimation techniques designed for hybrid architectures do not account for this coupling, leading to estimation inaccuracies due to the neglect of radiation pattern effects on the measured channel \cite{A. Alkhateeb 2014 CE}, \cite{J. Lee}. To address this limitation, it is essential to refine traditional estimation methods.
Leveraging the inherent sparsity of the scattering cluster channel, we employ a compressed sensing (CS) algorithm for efficient channel estimation. Specifically, we construct a revised dictionary matrix that explicitly incorporates the impact of pattern-reconfigurable antennas, enabling a more accurate representation of the spatial-domain channel characteristics. This enhancement not only improves channel estimation accuracy but also facilitates a more robust and adaptive beamforming strategy, ultimately enhancing overall system performance.

Initially, the RF domain channel at each frame to be estimated is expressed as
\begin{equation}\label{eq:RF_channel}
\mathbf{h}_{\mathrm{RF},k}\triangleq\mathbf{F}_{\mathrm{EM}}^{H}\mathbf{H}_{\mathrm{EM},k}\mathbf{h}_{\mathrm{S},k},~\forall k,
\end{equation}
where $\mathbf{F}_{\mathrm{EM}}^{H}\mathbf{H}_{\mathrm{EM},k}$ reflects the radiation gains of the reconfigurable antennas based on the fixed radiation beamformer determined in the long-timescale, and it is apparent that this term is coupled with the spatial channel $\mathbf{h}_{\mathrm{S},k}$.
During the uplink training phase, we assume the $K$ UEs transmit pilot sequences over orthogonal time resources. Therefore, the BS can isolate each UE's signal, making channel estimation for different UEs completely independent. Suppose that $I$ pilot symbols are allocated to each UE. Taking the $k$-th UE as an example, the corresponding received signal during the $i$-th pilot at the BS $\mathbf{y}_{k}[i]\in\mathbb{C}^{N_{\mathrm{RF}}}$ can be expressed as
\begin{equation}
\begin{aligned}
\mathbf{y}_{k}[i]& = \mathbf{W}[i]^{H}\mathbf{h}_{\mathrm{RF},k}x[i]+\mathbf{W}[i]^{H}\mathbf{n}[i],~\forall k,i,
\end{aligned}
\end{equation}
where $\mathbf{W}[i]\triangleq\mathbf{F}_{\mathrm{RF}}[i]\mathbf{F}_{\mathrm{BB}}\in \mathbb{C}^{N_{\mathrm{t}}\times N_{\mathrm{RF}}}$ is the training combining matrix, $x[i]=1$ represents the transmit pilot, $\mathbf{n}[i]\sim\mathcal{CN}(\mathbf{0},\sigma^{2}\mathbf{I}_{N_{\mathrm{t}}})$ denotes the AWGN. The analog combining matrix $\mathbf{F}_{\mathrm{RF}}$ consists of constant-modulus entries randomly drawn from the quantized phase set $\mathcal{F}\triangleq \{\frac{1}{\sqrt{N_{\mathrm{t}}}}e^{\jmath\frac{2\pi b}{2^{B}}}|b=0, 1, \ldots, 2^{B}-1\}$ and the digital combining matrix is chosen as $\mathbf{F}_{\mathrm{BB}}\triangleq\mathbf{I}_{N_{\mathrm{RF}}}$ for simplicity. Collecting all $I$ pilot observations, the complete received pilot sequence at the BS for the $k$-th UE can be formulated as
\begin{equation}
\begin{aligned}
\label{CS_received signal}
\mathbf{y}_{k}
&=\mathbf{W}^{H}\mathbf{h}_{\mathrm{RF},k}+\mathbf{n},~\forall k,
\end{aligned}
\end{equation}
where
\begin{equation}
\begin{aligned}
  \mathbf{y}_{k} &\triangleq [\mathbf{y}_{k}[1]^{T},\mathbf{y}_{k}[2]^{T}, \ldots, \mathbf{y}_{k}[I]^{T}]^{T}\in \mathbb{C}^{N_{\mathrm{RF}}I}, \\
  \mathbf{W} &\triangleq [\mathbf{W}[1],\mathbf{W}[2], \ldots, \mathbf{W}[I]]\in \mathbb{C}^{N_{\mathrm{t}}\times N_{\mathrm{RF}}I}, \\
  \mathbf{n} &\triangleq [(\mathbf{W}[1]^{H}\mathbf{n}[1])^{T},(\mathbf{W}[2]^{H}\mathbf{n}[2])^{T}, \\
  &\quad\quad\quad\quad\quad\quad\ldots,(\mathbf{W}[I]^{H}\mathbf{n}[I])^{T}]^{T}\in \mathbb{C}^{N_{\mathrm{RF}}I}.
\end{aligned}
\end{equation}

\vspace{-0.3cm}
Considering that the number of RF chains is limited and excessive training overhead must be avoided, the dimension of the received pilot sequence is typically much smaller than the number of BS antennas, i.e., $N_{\mathrm{RF}}I \ll N_{\mathrm{t}}$. As a result, directly estimating the high-dimensional RF domain channel $\mathbf{h}_{\mathrm{RF},k}$ from the low dimensional observation $\mathbf{y}_{k}$ becomes highly challenging. Fortunately, the scattering cluster channel exhibits inherent sparsity in the angular domain. Leveraging this property, we adopt a CS–based channel estimation approach, under which the received signal in \eqref{CS_received signal} can be equivalently represented as
\begin{equation}
\begin{aligned}
\mathbf{y}_{k}&=\mathbf{W}^{H}\mathbf{A}_{\mathrm{D}}\mathbf{z}_{k}+\mathbf{n}\\
&=\mathbf{Q}\mathbf{z}_{k}+\mathbf{n},~\forall k,
\end{aligned}
\end{equation}
where $\mathbf{A}_{\mathrm{D}}\in \mathbb{C}^{N_{\mathrm{t}}\times M}$ is the refined dictionary matrix incorporating the effects of pattern-reconfigurable antennas, $\mathbf{z}_{k}\in \mathbb{C}^{M}$  is a sparse angular-domain coefficient vector whose nonzero entries correspond to spatial angles of the rays in the clusters, and $\mathbf{Q}\triangleq\mathbf{W}^{H}\mathbf{A}_{\mathrm{D}}$ is the sensing matrix, which is required to exhibit low mutual coherence and approximate Restricted Isometry Property (RIP) to enable accurate sparse recovery. Under such conditions, a CS-based method only requires the total number of effective measurements to exceed the number of propagation paths i.e., $N_{\mathrm{RF}}I \geq L$, to guarantee that $\mathbf{z}_{k}$ can be uniquely identified.

We subsequently construct a refined dictionary matrix that explicitly embeds the radiation gain of pattern-reconfigurable antennas, yielding a more precise characterization of the spatial-domain channel.
The conventional dictionary matrix consists of a set of steering vectors corresponding to uniformly distributed candidate angles, given by
\begin{equation}
   \mathbf{E}=[\mathbf{e}(\vartheta_{1}),\mathbf{e}(\vartheta_{2}),\ldots,\mathbf{e}(\vartheta_{M})]\in\mathbb{C}^{N_{\mathrm{t}}\times M},
\end{equation}
where each column $\mathbf{e}(\vartheta_{m})\in\mathbb{C}^{N_{\mathrm{t}}}$ represents a steering vector pointing toward the angle $\vartheta_m =-\frac{\pi}{2}+ \frac{\pi m}{M},~m = 1,2,\ldots, M$.
To incorporate the influence of the radiation pattern on the channel representation, we refine the dictionary matrix as
\begin{equation}\label{modified dictionary}
\mathbf{A}_{\mathrm{D}}(:,m) = \overline{\mathbf{F}}_{\mathrm{EM}}(:,m)\odot\mathbf{e}(\vartheta_{m}),~\forall m,
\end{equation}
where $\overline{\mathbf{F}}_{\mathrm{EM}}\triangleq[\mathbf{f}_{\mathrm{EM},1}, \mathbf{f}_{\mathrm{EM},2}, \ldots, \mathbf{f}_{\mathrm{EM},N_{\mathrm{t}}}]^{T}\in \mathbb{R}^{N_{\mathrm{t}}\times M}$ stores the radiation patterns of all antennas configured within the current super-frame, and $\overline{\mathbf{F}}_{\mathrm{EM}}(:,m)$ denotes the radiation gains of the $N_{\mathrm{t}}$ antennas in the direction of $\vartheta_{m}$. This formulation effectively integrates both radiation pattern effects and angular information, enabling a more accurate estimation of spatial channel characteristics. Moreover, to prevent directional bias and ensure uniform scaling across all angular atoms, each column of the refined dictionary is normalized as
\begin{equation}
\mathbf{A}_{\mathrm{D}}(:,m) =\frac{\mathbf{A}_{\mathrm{D}}(:,m)}{\|\mathbf{A}_{\mathrm{D}}(:,m)\|},~\forall m.
\end{equation}

\vspace{-0.3cm}
Subsequently, the channel estimation problem can be formulated as a sparse recovery problem:
\begin{subequations}\label{eq:CS problem}
\begin{align}
    \min_{\mathbf{z}_{k}}&\ \ \
    \|\mathbf{y}_{k}-\mathbf{Q}\mathbf{z}_{k}\|_{2}\\
     \mathrm{s.t.}
    &\ \ \ \|\mathbf{z}_{k}\|_{0}=L.
\end{align}
\end{subequations}
This problem is typically solved using the orthogonal matching pursuit (OMP) algorithm, which iteratively selects the most relevant dictionary elements to approximate the channel. The iterations continue until the residual error falls below a predefined threshold or the sparsity constraint is met. The implementation details of the OMP algorithm and the design of the training hybrid beamformer follow the principles outlined in \cite{A. Alkhateeb 2014 CE}, \cite{J. Lee}. For brevity, these details are omitted here.

\subsubsection{Medium-timescale Analog Beamformer Design}
With the radiation beamformer $\mathbf{F}_{\mathrm{EM}}$ determined in the long-timescale, we now focus on designing the analog beamformer in the medium-timescale based on the statistical CSI of multi-path rays. Given the statistical information of the channel, the objective function is adjusted to maximize the ergodic sum-rate, ensuring robust performance over average channel conditions. Thus, the  analog beamformer design problem in the medium-timescale is formulated as follows 
\begin{subequations}
\label{eq:original_problem_RF}
\begin{align}
\label{eq:original_problem_RF_a}
    \max_{\mathbf{F}_{\mathrm{RF}}}&\ \ \sum^{K}_{k=1}\mathbb{E}_{\mathbf{H}_{\mathrm{RF}}}\{R_{k}\}\\
     \mathrm{s.t.}
    \label{eq:original_problem_RF_b}
    &\ \ \mathbf{F}_{\mathrm{RF}}(n,l_{\mathrm{RF}})\in \{\mathcal{F},0\}, ~\forall n,l_{\mathrm{RF}},\\
    \label{eq:original_problem_RF_c}
    &\ \ \|\mathbf{F}_{\mathrm{RF}}(n,:)\|_{0}=1, ~\forall n,
\end{align}
\end{subequations}
where the objective function \eqref{eq:original_problem_RF_a} defines the ergodic sum-rate evaluated over the RF domain channel samples, denoted as  $\mathbf{H}_{\mathrm{RF}}\triangleq[\mathbf{h}_{\mathrm{RF},1}, \mathbf{h}_{\mathrm{RF},2}, \ldots, \mathbf{h}_{\mathrm{RF},K}]\in \mathbb{C}^{N_{\mathrm{t}}\times K}$ with $\mathbf{h}_{\mathrm{RF},k}$ given in \eqref{eq:RF_channel}.
Notably, in this RF domain analog beamformer design problem, the power constraint is not explicitly imposed since the digital beamformer can be subsequently adjusted to satisfy it.

Unlike conventional statistical CSI-based beamforming approaches that require explicit knowledge of the channel statistics, we adopt an online learning framework that incrementally updates the channel statistical information by observing one channel sample at a time. This learning-based approach significantly reduces measurement overhead and latency, making it particularly well-suited for the tri-hybrid beamforming architecture. However, solving \eqref{eq:original_problem_RF} is challenging due to the combination of discrete phase constraints in  \eqref{eq:original_problem_RF_b} and the sparsity constraint in \eqref{eq:original_problem_RF_c}, which collectively impose a highly non-convex set for $\mathbf{F}_{\mathrm{RF}}$. To overcome these challenges, we decompose the analog beamformer into two components:
\begin{equation}\label{eq:FRF_SRFFset}
\mathbf{F}_{\mathrm{RF}}\triangleq \mathbf{S}_{\mathrm{RF}}\mathbf{F}_{\mathrm{set}},
\end{equation}
where $\mathbf{S}_{\mathrm{RF}}\in\{0,1\}^{N_{\mathrm{t}}\times N_{\mathrm{RF}}2^{B}}$ is the RF domain phase-shift selection matrix and $\mathbf{F}_{\mathrm{set}}$ is defined as $\mathbf{F}_{\mathrm{set}}\triangleq\mathbf{I}_{N_\mathrm{RF}}\otimes \mathbf{f}_\mathrm{set}$
with $\mathbf{f}_{\mathrm{set}}$ containing all possible quantized phase-shifts, yielding
\begin{equation}
\mathbf{f}_{\mathrm{set}}\triangleq \frac{1}{\sqrt{N_{\mathrm{t}}}}\left[1, e^{-\jmath\frac{2\pi}{2^{B}}}, \ldots, e^{-\jmath\frac{2\pi(2^{B}-1)}{2^{B}}}\right]^{T}.
\end{equation}
The binary matrix $\mathbf{S}_{\mathrm{RF}}$ encodes phase-shift selections, where $\mathbf{S}_{\mathrm{RF}}(n,l_{\mathrm{RFB}})=1$ indicates that the $n$-th antenna is assigned to the $l_{\mathrm{RFB}}/2^{B}$-th RF chain with the corresponding phase-shift value given by $\mathbf{F}_{\mathrm{set}}(l_{\mathrm{RFB}},l_{\mathrm{RFB}}/2^{B})$.
With $\mathbf{F}_{\mathrm{set}}$ predefined, the analog  beamformer design problem \eqref{eq:original_problem_RF} can be equivalently formulated as
\begin{subequations}
\label{eq:original_problem_SRF}
\vspace{-0.1cm}
\begin{align}
\label{eq:original_problem_SRF_a}
    \max_{\mathbf{S}_{\mathrm{RF}}}&\ \ \ \sum^{K}_{k=1}\mathbb{E}_{\mathbf{H}_{\mathrm{RF}}}\{R_{k}\}\\
     \mathrm{s.t.}
     \label{eq:original_problem_SRF_b}
    &\ \ \ \mathbf{S}_{\mathrm{RF}}(n,l_{\mathrm{RFB}})\in \{0,1\}, ~\forall n,l_{\mathrm{RFB}},\\
    \label{eq:original_problem_SRF_c}
    &\ \ \ \|\mathbf{S}_{\mathrm{RF}}(n,:)\|_{1}=1, ~\forall n,
\end{align}
\vspace{-0.1cm}
\end{subequations}
where the sum-rate expression in \eqref{eq:original_problem_SRF_a} can be recast as
\begin{equation}
\!\!\!\!R_{k}\!\!=\!\log_2 \!\left(\!1\!\!+\!\frac{|\mathbf{h}_{\mathrm{RF},k}^{H}\mathbf{S}_{\mathrm{RF}}\mathbf{F}_{\mathrm{set}}\mathbf{f}_{\mathrm{BB},k}|^{2}}{\sum_{j=1,j\neq k}^{K}\!|\mathbf{h}_{\mathrm{RF},k}^{H}\mathbf{S}_{\mathrm{RF}}\mathbf{F}_{\mathrm{set}}\mathbf{f}_{\mathrm{BB},j}|^{2}\!\!+\!\!\sigma_{k}^{2}} \!\right)\!,\forall k.\!\!\!\!\\
\end{equation}
The primary challenges in solving this problem stem from the stochastic nature of the objective function \eqref{eq:original_problem_SRF_a}, which lacks a closed-form expression, and the non-smooth, non-convex Boolean constraint \eqref{eq:original_problem_SRF_b}. To address these issues, we employ a stochastic successive convex approximation (SSCA) framework \cite{A. Liu two-stage}, which recursively refines the beamformer based on observed channel realizations. The expectation in the objective function is approximated using a quadratic surrogate function, allowing for online adaptation without requiring explicit statistical knowledge of the channel. Furthermore, the Boolean constraint is handled using a penalty-based approach, similar to the method described in Section III-B.

First, we construct a recursive quadratic surrogate function for the objective function \eqref{eq:original_problem_SRF_a}, leveraging both current and past channel samples. This recursive approach enables the BS to gradually learn the statistical properties of the channel by continuously accumulating channel observations, thereby enhancing beamforming performance over time and ensuring convergence. For convenience in the following expressions, we define $g_{0}(\mathbf{s}_{\mathrm{RF}}; \mathbf{H}_{\mathrm{RF}})\triangleq R_{k}$, and introduce $\mathbf{s}_{\mathrm{RF}}\triangleq \mathrm{vec}(\mathbf{S}_{\mathrm{RF}})$.

At the $t_{\mathrm{M}}$-th recursion, based on the acquired channel sample $\mathbf{H}_{\mathrm{RF}}^{(t_{\mathrm{M}})}$ and the previous RF domain phase-shift selection matrix $\mathbf{s}_{\mathrm{RF}}^{(t_{\mathrm{M}}-1)}$, we can substitute \eqref{eq:original_problem_SRF_a} with a quadratic surrogate function as follows
\begin{equation}
\label{eq:surrogate}
   f^{(t_{\mathrm{M}})}\!(\mathbf{s}_{\mathrm{RF}}\!)\!=\!v_{\mathbf{s}_{\mathrm{RF}}}^{(t_{\mathrm{M}})}+ (\!\mathbf{v}_{\mathbf{s}_{\mathrm{RF}}}^{(t_{\mathrm{M}})}\!)^{T}\!(\mathbf{s}_{\mathrm{RF}}-\mathbf{s}_{\mathrm{RF}}^{(t_{\mathrm{M}}\!-\!1)})- \tau\|\mathbf{s}_{\mathrm{RF}}-\mathbf{s}_{\mathrm{RF}}^{(t_{\mathrm{M}}\!-\!1)}\!\|^{2},
\end{equation}
where $\tau>0$ is a constant ensuring strong convexity of $\tau\|\mathbf{s}_{\mathrm{RF}}-\mathbf{s}_{\mathrm{RF}}^{(t_{\mathrm{M}}-1)}\|^{2}$.
The term $v_{\mathbf{s}_{\mathrm{RF}}}^{(t_{\mathrm{M}})}$ serves as an approximation of the objective function value $\mathbb{E}_{\mathbf{H}_{\mathrm{RF}}}\{g_{0}(\mathbf{s}_{\mathrm{RF}}; \mathbf{H}_{\mathrm{RF}})\}$, which is updated recursively as
\begin{equation} v_{\mathbf{s}_{\mathrm{RF}}}^{(t_{\mathrm{M}})} = (1-\eta^{(t_{\mathrm{M}})})v_{\mathbf{s}_{\mathrm{RF}}}^{(t_{\mathrm{M}}-1)} + \eta^{(t_{\mathrm{M}})}g_{0}(\mathbf{s}_{\mathrm{RF}}^{(t_{\mathrm{M}}-1)}; \mathbf{H}_{\mathrm{RF}}^{(t_{\mathrm{M}})}),
\end{equation}
where $\eta^{(t_{\mathrm{M}})}$ is a properly chosen sequence to satisfy convergence requirements \cite{A. Liu two-stage}.
Similarly, the gradient term $\mathbf{v}_{\mathbf{s}_{\mathrm{RF}}}^{(t_{\mathrm{M}})}$ is an approximation of the partial derivative of $\mathbb{E}_{\mathbf{H}_{\mathrm{RF}}}\{g_{0}(\mathbf{s}_{\mathrm{RF}}; \mathbf{H}_{\mathrm{RF}})\}$ with respect to $\mathbf{s}_{\mathrm{RF}}$,  evaluated at $\mathbf{s}_{\mathrm{RF}}=\mathbf{s}_{\mathrm{RF}}^{(t_{\mathrm{M}}-1)}$. This is recursively computed as
\begin{equation}
\!\!\mathbf{v}_{\mathbf{s}_{\mathrm{RF}}}^{(t_{\mathrm{M}})} \!\!=\!\! (1\!-\!\eta^{(t_{\mathrm{M}})})\mathbf{v}_{\mathbf{s}_{\mathrm{RF}}}^{(t_{\mathrm{M}}\!-\!1)} \!+\! \eta^{(t_{\mathrm{M}})}\nabla_{\mathbf{s}_{\mathrm{RF}}}g_{0}(\mathbf{s}_{\mathrm{RF}}^{(t_{\mathrm{M}}\!-\!1)}; \mathbf{H}_{\mathrm{RF}}^{(t_{\mathrm{M}})}).
\end{equation}
The gradient of $g_{0}(\mathbf{s}_{\mathrm{RF}}; \mathbf{H}_{\mathrm{RF}})$ with respect to $\mathbf{s}_{\mathrm{RF}}$ is given by
\begin{equation}
\label{eq:gradient_s}
   \nabla_{\mathbf{s}_{\mathrm{RF}}}g_{0}(\mathbf{s}_{\mathrm{RF}}; \mathbf{H}_{\mathrm{RF}}) = \sum_{k=1}^{K}\left(\frac{\sum_{i}^{K}\mathbf{e}_{k,i}}{\mit{\Gamma_{k}}}-\frac{\sum_{i\neq k}^{K}\mathbf{e}_{k,i}}{\mit{\Gamma_{-k}}}\right),
\end{equation}
where we define
\begin{subequations}
\begin{align}
\mit{\Gamma_{k}} \!&=  \! \sum_{i}^{K}\left|\mathbf{h}_{\mathrm{RF},k}^{H}\mathbf{S}_{\mathrm{RF}}\mathbf{F}_{\mathrm{set}}\mathbf{f}_{\mathrm{BB},i}\right|^{2}+\sigma_{k}^{2}, ~\forall k, \\
\mit{\Gamma_{-k}}\! &=  \!\sum_{i\neq k}^{K}\left|\mathbf{h}_{\mathrm{RF},k}^{H}\mathbf{S}_{\mathrm{RF}}\mathbf{F}_{\mathrm{set}}\mathbf{f}_{\mathrm{BB},i}\right|^{2}+\sigma_{k}^{2}, ~\forall k,\\
\mathbf{e}_{k,i}\!  &= 2\Re\{\mathrm{vec}((\mathbf{h}_{\mathrm{RF},k}\mathbf{h}_{\mathrm{RF},k}^{H})\mathbf{S}_{\mathrm{RF}}(\mathbf{F}_{\mathrm{set}}\mathbf{f}_{\mathrm{BB},i}\mathbf{f}_{\mathrm{BB},i}^{H}\mathbf{F}_{\mathrm{set}}^{H}))\}. 
\end{align}
\end{subequations}
Due to space limitations, we omit the detailed derivations here. The surrogate function  $f^{(t_{\mathrm{M}})}(\mathbf{s}_{\mathrm{RF}})$ in \eqref{eq:surrogate} provides a concave approximation of the original stochastic objective function $\mathbb{E}_{\mathbf{H}_{\mathrm{RF}}}\{g_{0}(\mathbf{s}_{\mathrm{RF}}; \mathbf{H}_{\mathrm{RF}})\}$ in \eqref{eq:original_problem_SRF_a}, enabling efficient processing of problem \eqref{eq:original_problem_SRF} without explicitly computing the expectation.

To deal with the Boolean constraint \eqref{eq:original_problem_SRF_b}, we employ the same technique as specified in \eqref{eq:Boolean}-\eqref{eq:Boolean_MM_EM}. As a result, the analog beamformer design problem is formulated as
\begin{subequations}\label{eq:SRF_final}
\begin{align}
    \max_{\mathbf{S}_{\mathrm{RF}}}&\ \ \ f^{(t_{\mathrm{M}})}(\mathbf{s}_{\mathrm{RF}})+\varrho_{2}(2\mathbf{s}_{\mathrm{RF}}^{(t_{\mathrm{M}}-1)}-\mathbf{1})^{T}\mathbf{s}_{\mathrm{RF}}\\
     \mathrm{s.t.}
    &\ \ \ \mathbf{S}_{\mathrm{RF}}(n,l_{\mathrm{RFB}})\in [0,1], ~\forall n,l_{\mathrm{RFB}},\\
    &\ \ \ \|\mathbf{S}_{\mathrm{RF}}(n,:)\|_{1}=1, ~\forall n,
\end{align}
\end{subequations}
in which $\varrho_{2}$ denotes the penalty parameter. The resulting problem can be efficiently solved using a variety of existing methods or convex optimization solvers.
Once the optimal solution, denoted as $\mathbf{\overline{S}}^{(t_{\mathrm{M}})}_{\mathrm{RF}}$, is obtained, the phase-shift selection matrix $\mathbf{S}_{\mathrm{RF}}$ is updated to incorporate both the newly computed solution and the previous information. This update is performed using the following recursive expression:
\begin{equation}
\label{eq:update_SRF}
   \mathbf{S}_{\mathrm{RF}}^{(t_{\mathrm{M}})} = (1-\eta^{(t_{\mathrm{M}})})\mathbf{S}_{\mathrm{RF}}^{(t_{\mathrm{M}}-1)} + \eta^{(t_{\mathrm{M}})}\mathbf{\overline{S}}^{(t_{\mathrm{M}})}_{\mathrm{RF}}.
\end{equation}
With the updated phase-shift selection matrix $\mathbf{S}_{\mathrm{RF}}^{(t_{\mathrm{M}})}$, the analog beamformer is then constructed as $\mathbf{F}_{\mathrm{RF}}^{(t_{\mathrm{M}})}=\mathbf{S}_{\mathrm{RF}}^{(t_{\mathrm{M}})}\mathbf{F}_{\mathrm{set}}$.

\vspace{-0.2cm}
\subsection{Short-timescale Digital  Beamformer Design}
In the $t_{\mathrm{S}}$-th short time-slot of the $t_{\mathrm{M}}$-th frame, the digital beamformer is updated while keeping the radiation beamformer $\mathbf{F}_{\mathrm{EM}}$ and the analog beamformer $\mathbf{F}_{\mathrm{RF}}$ fixed. The update relies on the instantaneous low-dimensional effective CSI, denoted as $\mathbf{H}_{\mathrm{e}}^{(t_{\mathrm{S}})}=(\mathbf{H}_{\mathrm{RF}}^{(t_{\mathrm{S}})})^{H}\mathbf{F}_{\mathrm{RF}}^{(t_{\mathrm{M}})}\in \mathbb{C}^{K\times N_{\mathrm{RF}}}$, which is obtained from the processing of pilot signals in the BB domain. To design the digital beamformer, we apply the classic minimum mean square error (MMSE) method. The resulting digital beamformer is given by
\begin{equation}\label{eq:digital beamformer 2}
   \mathbf{\widetilde{F}}_{\mathrm{BB}}^{(t_{\mathrm{S}})} =(\mathbf{H}_{\mathrm{e}}^{(t_{\mathrm{S}})})^{H}\left( \mathbf{H}_{\mathrm{e}}^{(t_{\mathrm{S}})}(\mathbf{H}_{\mathrm{e}}^{(t_{\mathrm{S}})})^{H} + \bm{\Lambda}\right)^{-1},
\end{equation}
where $\bm{\Lambda}\triangleq\mathrm{diag}\{\sigma_{1}^{2},\sigma_{2}^{2}, \ldots, \sigma_{K}^{2}\}$.
Finally, to ensure compliance with the total power constraint, the digital beamformer is normalized as
\begin{equation}
    \label{eq:digital beamformer 2_power}    \mathbf{F}_{\mathrm{BB}}^{(t_{\mathrm{S}})}=\frac{\sqrt{P_{\mathrm{t}}}\ \mathbf{\widetilde{F}}_{\mathrm{BB}}^{(t_{\mathrm{S}})}}{\|\mathbf{F}_{\mathrm{RF}}^{(t_{\mathrm{M}})}\mathbf{\widetilde{F}}_{\mathrm{BB}}^{(t_{\mathrm{S}})}\|_{F}}.
\end{equation}

\begin{algorithm}[t]
\caption{The proposed tri-timescale tri-hybrid beamforming design algorithm.}
\label{alg:3}
\begin{algorithmic}[0]
\begin{small}
    \STATE {\!\!\!\!\!\!\textbf{Initialize} $\mathbf{F}_{\mathrm{EM}}, \mathbf{F}_{\mathrm{RF}}, \mathbf{F}_{\mathrm{BB}}$.}
    \STATE {\!\!\!\!\!\!\textbf{Step 1: Long-timescale EM domain design at the $t_{\mathrm{L}}$-th super-frame} }
        \begin{itemize}[leftmargin=0.35cm, itemindent=0cm]
        \item  High-dimensional scattering cluster core channel estimation
            \begin{itemize}[leftmargin=0.4cm, itemindent=0cm]
            \item Calculate the channel covariance matrix $\mathbf{R}_{k}^{(t_{\mathrm{L}})}, \forall k$ by \eqref{eq:channel correlated matrix}.
            \item Derive the APS $\rho^{(t_{\mathrm{L}})}_{k}(\vartheta_{m}), \forall k$ as \eqref{eq:APS}.
            \item Estimate the nominal angles $\theta_{k,c}, \forall k,c,$ and average channel gains $\alpha_{k,c}, \forall k,c,$ of the scattering clusters.
            \item Construct the EM domain channel $\mathbf{h}^{(t_{\mathrm{L}})}_{\mathrm{C},k}$ by \eqref{eq:EM_channel}.
            \end{itemize}
        \item  Radiation beamformer design
            \begin{itemize}[leftmargin=0.4cm, itemindent=0cm]
            \item Calculate EM domain pattern selection matrix $\mathbf{S}_{\mathrm{EM}}$.\\
            \begin{itemize}[leftmargin=0.3cm ,label={},]
                \WHILE {no convergence of $\mathbf{S}_{\mathrm{EM}}$}
                   \STATE {Update $\bm{\mu}$ by \eqref{eq:mu}.}
                   \STATE {Update $\bm{\xi}$ by \eqref{eq:xi}.}
                   \STATE {Update pattern selection matrix $\mathbf{S}_{\mathrm{EM}}$ by solving \eqref{eq:EM problem final}.}
                \ENDWHILE
            \end{itemize}
            \item Construct the radiation beamformer $\mathbf{F}^{(t_{\mathrm{L}})}_{\mathrm{EM}}$ by \eqref{eq:FEM_decomposition}.
                  \end{itemize}
        \end{itemize}
        \STATE {\!\!\!\!\!\!\textbf{Step 2: Medium-timescale RF domain design at the $t_{\mathrm{M}}$-th frame} }
            \begin{itemize}[leftmargin=0.35cm, itemindent=0cm]
            \item  Statistical high-dimensional multi-path rays channel estimation
            \begin{itemize}[leftmargin=0.4cm, itemindent=0cm]
            \item Construct the modified dictionary matrix $\mathbf{A}^{(t_{\mathrm{M}})}_{\mathrm{D}}$ by \eqref{modified dictionary}.
            \item Solve the channel estimation problem \eqref{eq:CS problem} by the OMP algorithm.
            \item Construct the RF domain channel $\mathbf{h}^{(t_{\mathrm{M}})}_{\mathrm{RF},k}, \forall k,$ by \eqref{eq:RF_channel}.
            \end{itemize}
            \item  Analog beamformer design
            \begin{itemize}[leftmargin=0.4cm, itemindent=0cm]
            \item Update the surrogate function \eqref{eq:surrogate} based on $\mathbf{H}_{\mathrm{RF}}^{(t_{\mathrm{M}})}, \mathbf{S}_{\mathrm{RF}}^{(t_{\mathrm{M}}-1)}$.
            \item Calculate the optimal $\mathbf{\overline{S}}_{\mathrm{RF}}^{(t_{\mathrm{M}})}$ using \eqref{eq:SRF_final}.
            \item Update $\mathbf{S}_{\mathrm{RF}}^{(t_{\mathrm{M}})}$ according to \eqref{eq:update_SRF}.
            \item Construct the analog beamformer $\mathbf{F}_{\mathrm{RF}}^{(t_{\mathrm{M}})}$ by \eqref{eq:FRF_SRFFset}.
            \end{itemize}
            \end{itemize}
        \STATE {\!\!\!\!\!\!\textbf{Step 3: Short-timescale BB domain design at the $t_{\mathrm{S}}$-th time-slot}}
            \begin{itemize}[leftmargin=0.35cm, itemindent=0cm]
            \item Instantaneous low-dimensional effective channel estimation
            \begin{itemize}[leftmargin=0.4cm, itemindent=0cm]
            \item Observe an instantaneous low-dimensional effective channel sample $\mathbf{H}_{\mathrm{e}}^{(t_{\mathrm{S}})}$.
            \end{itemize}
            \item Digital beamformer design
            \begin{itemize}[leftmargin=0.4cm, itemindent=0cm]
            \item With fixed $\mathbf{F}_{\mathrm{EM}}^{(t_{\mathrm{L}})}$ and $\mathbf{F}_{\mathrm{RF}}^{(t_{\mathrm{M}})}$, calculate the digital beamformer $\mathbf{F}_{\mathrm{BB}}^{(t_{\mathrm{S}})}$ according to \eqref{eq:digital beamformer 2} and \eqref{eq:digital beamformer 2_power}.
            \end{itemize}
            \end{itemize}
    \end{small}
    \end{algorithmic}
\end{algorithm}

\subsection{Summary, Complexity and  Overhead Analysis}
\subsubsection{Summary}
Based on the above derivations, the proposed tri-timescale tri-hybrid beamforming design is straightforward and summarized in Algorithm 1. After appropriately initializing the beamforming matrices $\mathbf{F}_{\mathrm{EM}}$,  $\mathbf{F}_{\mathrm{RF}}$ and  $\mathbf{F}_{\mathrm{BB}}$ across the three domains, the algorithm alternately acquires the corresponding channel information and updates these matrices over the three timescales, respectively.

\subsubsection{Complexity Analysis}
Next, we provide a brief analysis of the algorithmic complexity. The primary computational complexity lies in calculating the EM domain pattern selection matrix $\mathbf{S}_{\mathrm{EM}}$ in \eqref{eq:EM problem final} over the long-timescale, and determining the  phase-shift selection matrix $\mathbf{S}_{\mathrm{RF}}$ in \eqref{eq:SRF_final} over the medium-timescale.
Specifically, the computational complexity of solving problem \eqref{eq:EM problem final} in each super-frame over the long-timescale with an $N_{\mathrm{t}}P$-dimensional variable, is of order $\mathcal{O}\{N_{\mathrm{t}}^{3.5}P^{3.5}\}$ \cite{Lectures 2001}. Similarly, the computational complexity of solving problem \eqref{eq:SRF_final} in each frame during the medium-timescale with an $N_{\mathrm{t}}N_{\mathrm{RF}}2^{B}$-dimensional variable  is of order $\mathcal{O}\{N_{\mathrm{t}}^{3.5}N_{\mathrm{RF}}^{3.5}2^{3.5B}\}$.

\subsubsection{Pilot Overhead Analysis}
The pilot overhead in the proposed scheme primarily stems from channel sampling at the medium-timescale and short-timescale stages. In the medium-timescale, CS-based channel estimation requires $I$ pilot symbols $(N_{\mathrm{RF}}I\gg L)$ for each UE to obtain the RF domain high-dimensional statistical channel sample $\mathbf{h}_{\mathrm{RF},k}^{(t_{\mathrm{M}})}$, as described in Section III-C. In the short-timescale, $N_{\mathrm{RF}}K$ pilot symbols are needed for estimating the BB domain low-dimensional effective channel $\mathbf{H}_{\mathrm{e}}^{(t_{\mathrm{S}})}$, as described in Section III-D.
Therefore, the total pilot overhead across the channel statistics coherence time is expressed as $(KI+KT_{\mathrm{S}})T_{\mathrm{M}}T_{\mathrm{L}}$. This is notably less than that of the real-timescale beamforming approaches, which requires $(KI+K)T_{\mathrm{S}}T_{\mathrm{M}}T_{\mathrm{L}}$.

\subsubsection{Initialization}
During the initialization phase, we set all antenna patterns to point toward the boresight direction for the radiation beamformer $\mathbf{F}_{\mathrm{EM}}$, adopt a random phase initialization for the analog beamformer $\mathbf{F}_{\mathrm{RF}}$, and employ a normalized identity matrix for the digital beamformer $\mathbf{F}_{\mathrm{BB}}$. These initial settings serve as a neutral starting point for the subsequent optimization.

\vspace{-0.3cm}
\section{Simulation Results}
In this section, we provide simulation results to demonstrate the effectiveness of the proposed tri-timescale tri-hybrid beamforming design algorithm with reconfigurable antennas in enhancing communication performance while ensuring hardware efficiency.
Unless otherwise specified, the BS is equipped with $N_{\mathrm{t}} = 32$ antennas and the center carrier frequency is set to $f_{\mathrm{c}}= 28$ GHz. The low-resolution PS used in the RF domain has quantized phase controlled by $B=3$ bits. For simplicity and to focus on the tri-timescale design, we set $N_{\mathrm{RF}}=K$ in simulations and the framework applies to $N_{\mathrm{RF}}\geq K$ as well. In the EM domain, each pixel-based reconfigurable antenna enables the generation of $P=7$ distinct directional radiation patterns by adjusting the electronic switches between pixels \cite{R. Murch 2022 pixel pattern endfire}, with $M=180$ uniform sampling points in the azimuth dimension\footnote{The prototype reconfigurable antenna in \cite{R. Murch 2022 pixel pattern endfire} currently has an aperture exceeding the conventional half-wavelength spacing, introducing practical challenges such as mutual coupling and grating lobes, which we leave for future investigation. In this study, we adopt half-wavelength spacing to facilitate fair comparisons with conventional array designs. With ongoing advancements in reconfigurable antenna technology, recent studies and prototypes have convincingly demonstrated the feasibility of smaller apertures compatible with half-wavelength spacing, thereby addressing these practical concerns.}.
In this scenario, there are $K=2$ UEs, all operating under a scattering cluster channel with $C=2$ clusters, each of which consists of $L_{c}=3$ multi-path rays. The nominal AoDs $\overline{\theta}_{k,c}$ of these scattering clusters are randomly distributed within $[-\pi/2,\pi/2]$, while the angle spread is set to $\varsigma_{k,c}=\pi/36$. The large-scale path loss is expressed as $\alpha_{k,c}=10^{-\frac{C_{0}}{10}}(\frac{r_{k,c}}{D_{0}})^{-\kappa_{k,c}}$ with $\kappa_{k,c}\in[2.5, 3]$ being the attenuation factor, $C_{0}=30$ dB, $D_{0}=1$m. The propagation distance $r_{k,c}$ of each scattering cluster is randomly distributed in $r_{k,c}\in[50\mathrm{m},100\mathrm{m}]$. The noise power of UEs is set as $\sigma_{k}^{2}=-70$dBm.
The tri-timescale framework is structured with $T_{\mathrm{L}}=10$ super-frames for the long-timescale, $T_{\mathrm{M}}=50$ frames for the medium-timescale, and $T_{\mathrm{S}}=200$ time-slots for the short-timescale.

\begin{figure}[t]
    \centering
    \subfigure[Long-timescale.]{\includegraphics[width= 1.65 in]{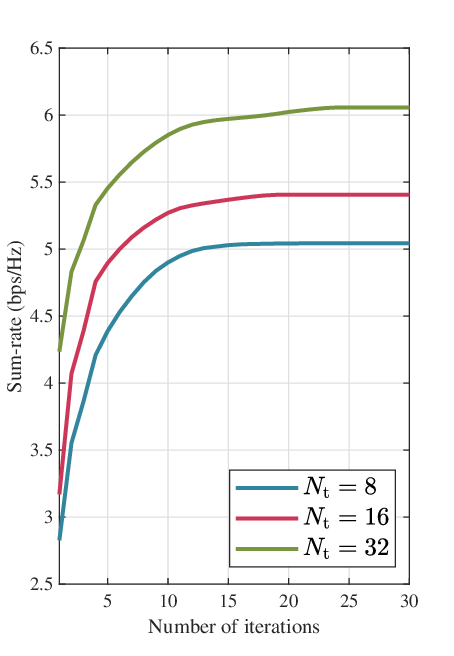}}%
    \vspace{-0.0 cm}
    \subfigure[Medium-timescale.]{\includegraphics[width= 1.65 in]{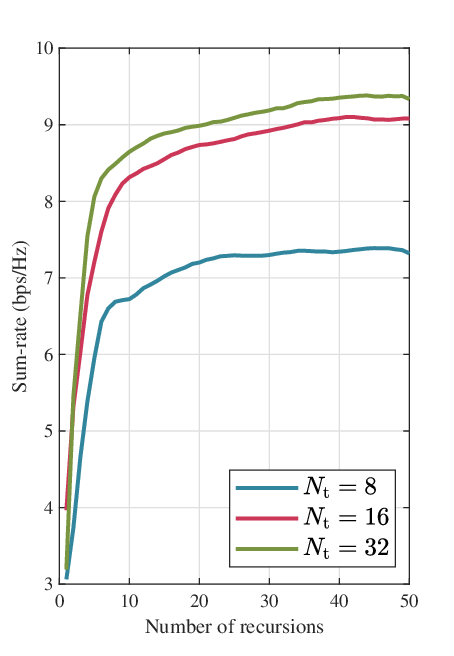}}
    \caption{Convergence performance of the radiation  beamformer design and analog beamformer design.}
     \vspace{-0.3 cm}
    \label{fig:iteration}
\end{figure}

\begin{figure}[!t]
\vspace{-0.2cm}
  \centering
  \includegraphics[width= 3.3 in]{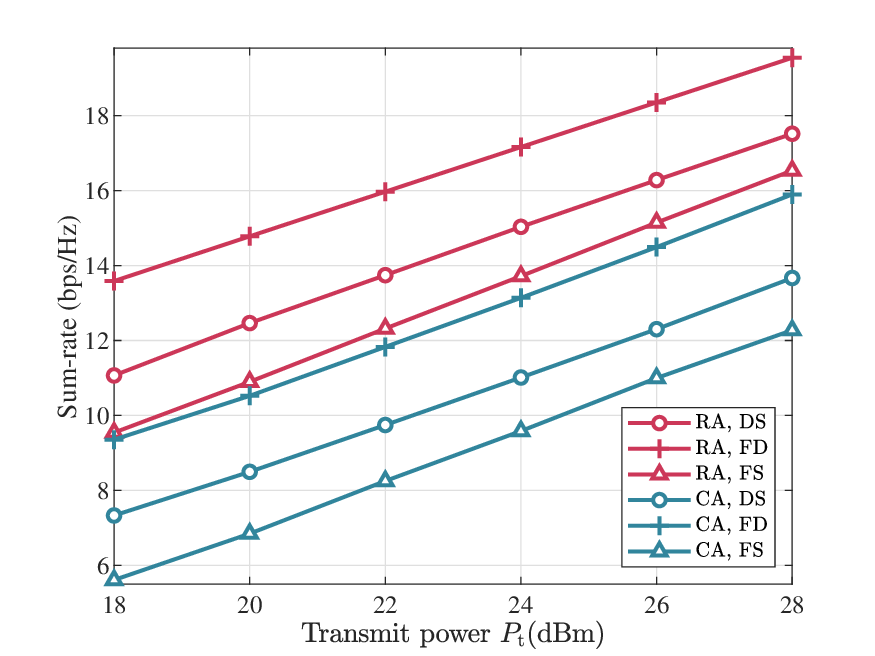}
  \caption{The sum-rate versus the transmit power $P_\mathrm{t}$.}
  \label{fig:P}
   \vspace{-0.3 cm}
\end{figure}

The convergence performance of the proposed tri-timescale tri-hybrid beamforming design is illustrated in Fig. \ref{fig:iteration}, which includes both the long-timescale radiation beamformer design algorithm and the medium-timescale analog beamformer design algorithm.
It should be emphasized that the convergence implications differ between the two algorithms. In the long-timescale case, convergence refers to the iterative calculation of the radiation beamformer at the start of super-frame, based on the estimated CSI of the scattering cluster core. In contrast, for the medium-timescale case, the convergence process refers to the recursive learning of the channel statistics of the multi-path rays within the scattering clusters using high-dimensional channel samples and the recursive updating of the analog beamformer in each frame. The results presented in Fig. \ref{fig:iteration} demonstrate that both algorithms exhibit satisfactory convergence speed, indicating an efficient level of computational complexity.

\begin{figure}[!t]
    \centering
    \vspace{0.0cm}
    \subfigure[]{\includegraphics[width= 1.65 in]{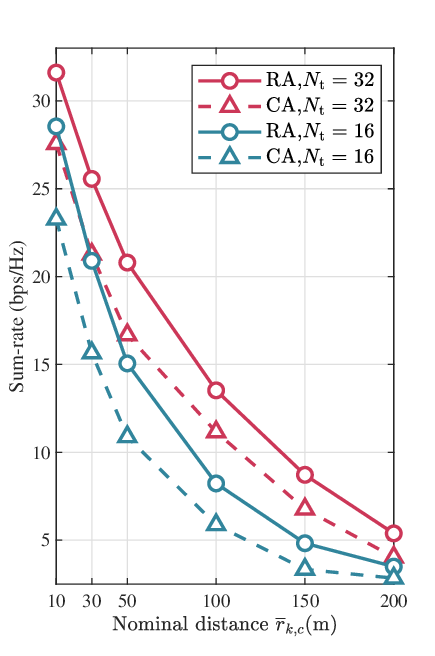}}%
    \vspace{-0.0 cm}
    \subfigure[]{\includegraphics[width= 1.65 in]{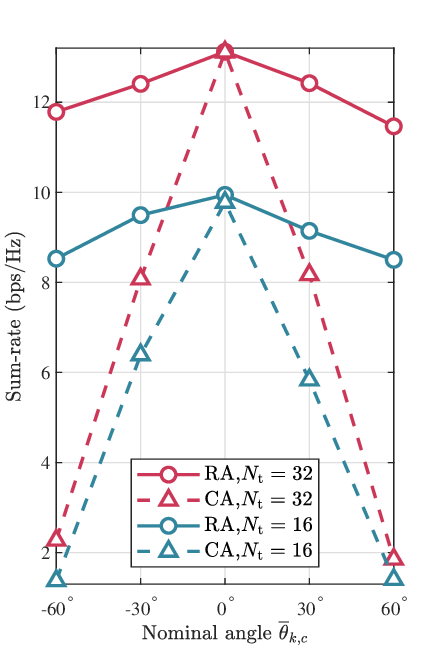}}
    \caption{The sum-rate versus the distance and angle of UE.}
    \vspace{-0.3 cm}
    \label{fig:Rtheta}
\end{figure}

\begin{figure}[!t]
  \centering
  \includegraphics[width= 3.3 in]{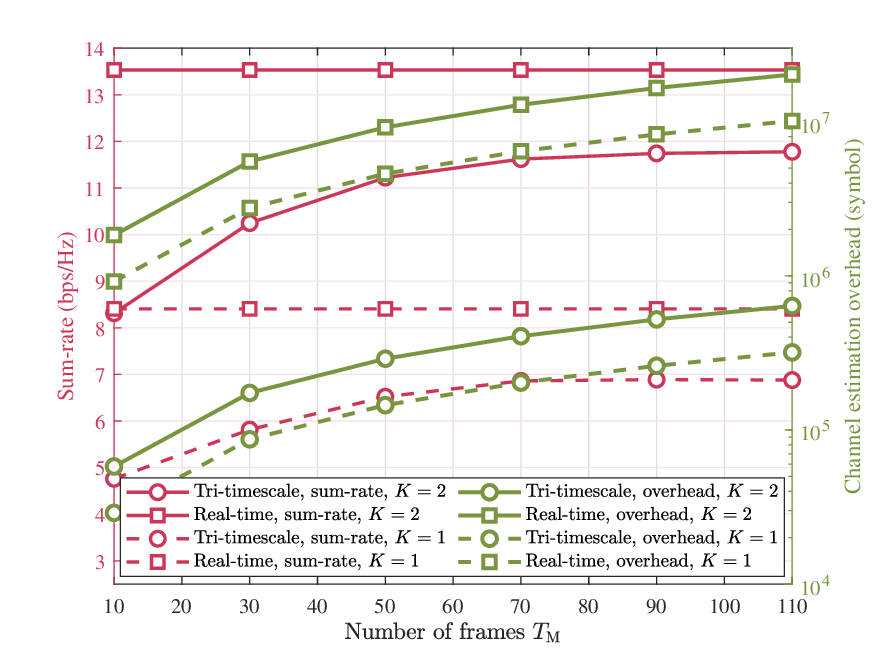}
  \caption{The sum-rate and channel estimation overhead versus the number of frames $T_{\mathrm{M}}$.}
  \label{fig:T}
  \vspace{-0.3 cm}
\end{figure}

Fig. \ref{fig:P} illustrates the sum-rate performance versus transmit power for systems employing reconfigurable antennas (RA) and conventional antennas (CA). The proposed tri-hybrid beamforming architecture with reconfigurable antennas and dynamic-subarray selection is labeled as ``\textbf{RA, DS}''. For benchmarking purposes, fully-digital beamforming with reconfigurable antennas denoted as ``\textbf{RA, FD}'' serves as the upper performance bound and corresponds to a real-timescale optimization, while hybrid beamforming with fixed-subarray selection denoted as``\textbf{RA, FS}'' represents the lower bound. Specifically, the ``\textbf{RA, DS} and ``\textbf{RA, FS}'' configurations implement the proposed tri-timescale optimization framework, where EM domain, RF domain, and BB domain beamformers are optimized across distinct timescales.
For comparison, we also include three schemes employing conventional antennas: ``\textbf{CA, DS}'', ``\textbf{CA, FD}'', and ``\textbf{CA, FS}''. Among them, “\textbf{CA, FD}” corresponds to a real-timescale design, whereas ``\textbf{CA, DS}'' and ``\textbf{CA, FS}'' employ a conventional two-timescale optimization approach, optimizing analog beamformers over a long timescale and digital beamformers over a short timescale \cite{A. Liu two-stage}-\cite{A. Liu 2022 sparse}.
As observed in Fig. \ref{fig:P}, the proposed ``\textbf{RA, DS}'' approach consistently outperforms other schemes. In particular, the use of reconfigurable antennas achieves approximately a 7 dB gain compared to conventional antennas. Furthermore, employing dynamic-subarray selection yields an additional gain of roughly 2 dB over fixed-subarray schemes. These results validate that the proposed tri-hybrid beamforming architecture, benefiting from adaptive EM domain radiation patterns and flexible RF domain subarray selections, significantly enhances overall communication performance.


\begin{figure}[!t]
    \centering
    \vspace{0.0cm}
    \subfigure[Long-timescale.]{\includegraphics[width= 3.2 in]{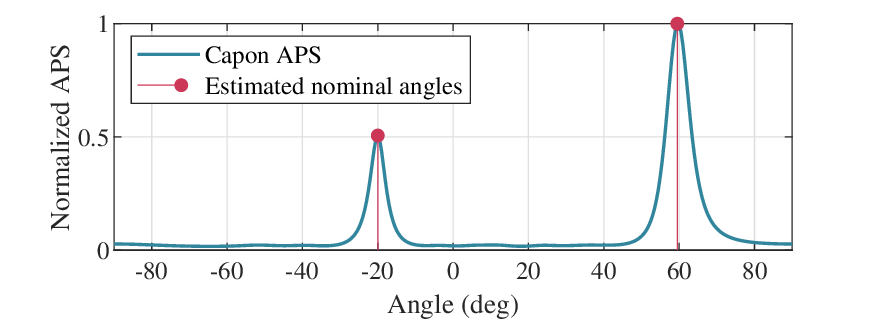}}%
    \vspace{-0.1 cm}
    \subfigure[Medium-timescale.]{\includegraphics[width= 3.2 in]{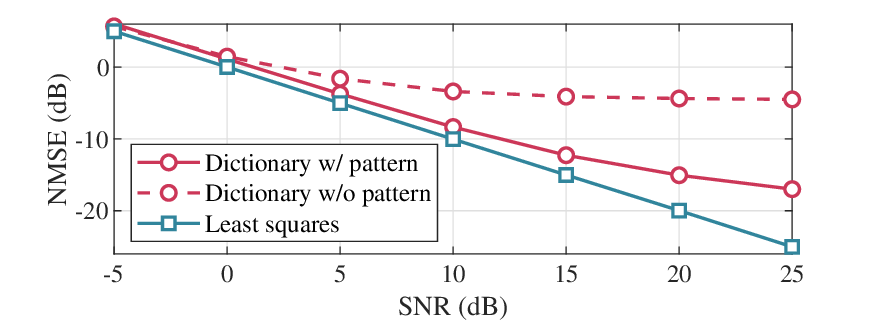}}
    \vspace{-0.2 cm}
    \caption{The performance evaluation of channel estimation in the long-timescale and medium-timescale.}
    \vspace{-0.3 cm}
    \label{fig:CSI}
\end{figure}

\begin{figure}[!t]
  \centering
  \includegraphics[width= 3.3 in]{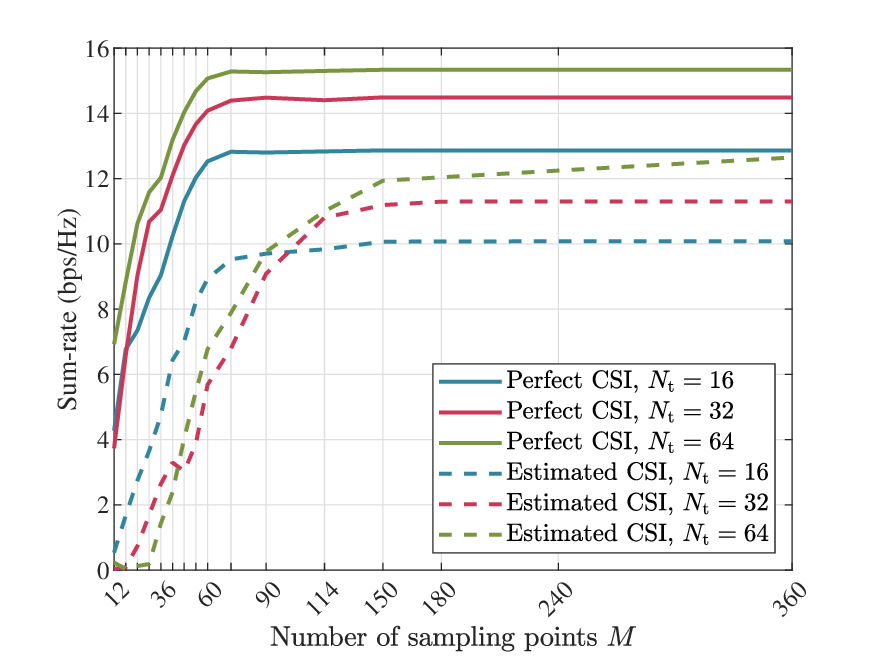}
  \caption{The sum-rate versus the number of angular sampling points $M$.}
  \label{fig:M}
   \vspace{-0.3 cm}
\end{figure}

For the proposed tri-timescale tri-hybrid beamforming design, increasing the channel sampling rate in the medium-timescale can enhance sum-rate performance by enabling more frequent updates of channel statistics for multi-path rays and scattering cluster cores. However, this improvement comes at the cost of higher pilot overhead. To quantitatively evaluate the impact of the channel sampling rate on both system performance and pilot overhead, we gradually increase the number of frames per super-frame in the medium-timescale,  effectively increasing the channel sampling rate, as illustrated in Fig. \ref{fig:T}. By analyzing the pilot overhead expression derived in Section III-E and observing the trend in Fig. \ref{fig:T}, we conclude that the proposed tri-timescale scheme significantly reduces pilot overhead compared to the real-time scheme, which requires high-dimensional CSI estimation in every time-slot. More importantly, as the number of frames increases, the sum-rate performance of the tri-timescale beamforming scheme steadily improves and eventually saturates at a satisfactory level, albeit with a monotonic increase in channel estimation overhead. In particular, selecting the number of frames as $T_{\mathrm{M}}=50$ achieves an optimal balance between performance and overhead, and this configuration is therefore employed in our simulations.


\begin{figure}[!t]
    \centering
    \vspace{0.0cm}
    \subfigure[]{\includegraphics[width= 1.65 in]{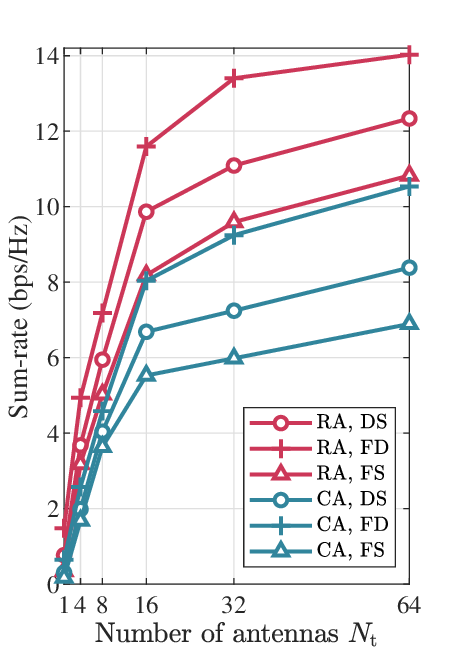}}%
    \vspace{-0.0 cm}
    \subfigure[]{\includegraphics[width= 1.65 in]{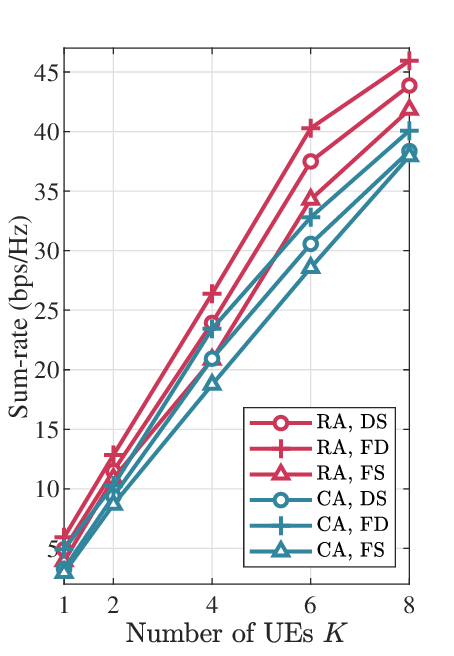}}
    \caption{The sum-rate versus the number of antennas and UEs.}
    \vspace{-0.3 cm}
    \label{fig:NK}
\end{figure}

Next, we validate the effectiveness of the proposed scheme by evaluating the impact of user location on the overall performance of the tri-hybrid beamforming with reconfigurable antennas and the  hybrid beamforming with conventional antennas. Specifically, Fig. \ref{fig:Rtheta} (a) depicts the sum-rate performance as a function of the nominal distance, which represents the distance between the BS and the UEs, while Fig. \ref{fig:Rtheta} (b) illustrates the sum-rate variation with respect to the nominal angle, which represents the user's angular position relative to the antenna array.
From Fig. \ref{fig:Rtheta} (a), we observe that as the nominal distance increases,
the performance gain offered by pattern-reconfigurable antennas gradually diminishes. This is primarily due to the dominant effect of path loss, which attenuates signal power, reducing the relative impact of radiation pattern variations across different angles. In contrast, Fig. \ref{fig:Rtheta} (b) reveals that the RA-based scheme maintains robust performance across a wide range of user angles by dynamically adjusting the pattern orientations as the user's position deviates from the optimal direction. In comparison, the CA-based scheme achieves satisfactory performance only when the user is positioned perpendicular to the array, where the antenna gain is maximized. These results underscore the robustness and versatility of the proposed tri-hybrid beamforming approach, demonstrating that dynamic pattern reconfiguration not only compensates for variations in user location but also substantially enhances overall system performance and reliability across a wide range of deployment scenarios.

The channel estimation algorithms in both the EM and RF domains are critical components of the proposed tri-timescale framework. As shown in Fig. \ref{fig:CSI} (a),
the normalized APS obtained by the Capon method is depicted as a function of the angle. Two dominant peaks can be clearly observed, corresponding to the scattering clusters in the angular domain. The estimated nominal angles, marked by red dots, are well aligned with the peaks of the Capon APS, indicating accurate cluster angle estimation in the long timescale.
The normalized mean square error (NMSE) performance of channel estimation is evaluated versus SNR for different schemes. The CS-based channel estimation using dictionary with pattern information consistently outperforms its counterpart without pattern information, demonstrating the benefit of exploiting pattern-aware dictionaries in the medium-timescale channel estimation process.

In Fig. 10, we investigate how the number of angular sampling points $M$ affects the achievable sum-rate. It is evident that the system sum-rate monotonically increases with $M$, since a finer angular grid provides a more accurate discretization of continuous radiation patterns, thus improving the matching precision between radiation gains and user angles. Nevertheless, the incremental benefit gradually becomes negligible as
$M$ increases beyond a certain point (around
$M=60$ for perfect CSI and $M=180$ for estimated CSI), indicating limited value in excessively dense angular sampling. Specifically, after comprehensive consideration, the choice of
$M=180$, adopted in earlier simulations, achieves a favorable balance between performance improvement and computational complexity, effectively capturing most performance gains without incurring excessive complexity.

To further examine the scalability of the proposed tri-hybrid beamforming architecture, Fig. 11 illustrates the achievable sum-rate performance under varying numbers of BS antennas and UEs. The results in Fig. 11 (a) clearly underscore the hardware efficiency and substantial array-size reductions achievable with reconfigurable antennas, highlighting their considerable potential for supporting energy-efficient next-generation wireless systems without compromising communication performance. Additionally, as depicted in Fig. 11 (b), the proposed tri-hybrid beamforming consistently outperforms both baseline hybrid beamforming and conventional antenna schemes across all considered user loads, demonstrating robust scalability. These comprehensive simulation results thus strongly validate the practicality and scalability of the proposed tri-hybrid architecture for future large-scale multi-user wireless deployments.

\vspace{-0.0 cm}
\section{Conclusions}
This paper presents a novel tri-hybrid beamforming architecture that integrates pattern-reconfigurable antennas with analog and digital beamforming. By leveraging an innovative tri-timescale beamforming framework, the proposed approach optimizes the radiation beamformer over a long-timescale, the analog beamformer over a medium-timescale, and the digital beamformer over a short-timescale. This hierarchical design effectively balances system performance with practical constraints, such as channel estimation overhead and computational complexity. Numerical results confirmed that the proposed design achieves around 7 dB sum-rate improvement, while reducing pilot overhead by nearly an order of magnitude compared with real-time joint beamforming schemes. The demonstrated scalability and performance gains underscore the practical feasibility and effectiveness of our approach for future large-scale wireless systems. Future work will address practical hardware aspects and extend this framework to wideband and near-field scenarios.

Future research directions include addressing practical hardware challenges associated with reconfigurable antennas, such as severe mutual coupling effects resulting from half-wavelength spacing. Incorporating multi-port circuit theory to model and mitigate mutual coupling through joint radiation pattern and hybrid beamforming design is a promising pathway. Furthermore,  larger antenna spacing can introduce strong grating lobes in practice. To address this issue, it will be essential to explore sparse or non-uniform array configurations, irregular sampling methods, and integrated radiation pattern optimization schemes.  Investigating these aspects will further extend the practical applicability and deepen the theoretical understanding of tri-hybrid beamforming architectures for realistic deployment scenarios.

\end{document}